\documentclass[aip,jcp,reprint]{revtex4-1}

\usepackage{graphicx} 
\usepackage{natbib}
\usepackage[utf8]{inputenc}
\usepackage{dcolumn}
\usepackage{amsmath}
\usepackage{amsfonts}
\usepackage{amssymb}
\usepackage[T1]{fontenc}
\usepackage{mathrsfs}
\usepackage{braket}
\usepackage{geometry}
\usepackage{here}
\usepackage{lmodern}
\usepackage{array}
\usepackage{url}
\usepackage{textcase}
\usepackage{bm}

\usepackage{xcolor}

\usepackage{multirow}
\usepackage{booktabs}

\begin{document}


\title{Managing Temperature in Open Quantum Systems Strongly Coupled with Structured Environments} 



\author{Brieuc Le D\'e}
 \affiliation{Institut des Nanosciences de Paris, Sorbonne Université, CNRS, F-75005 Paris, France} 

 \author{Amine Jaouadi}
 \affiliation{LyRIDS, ECE Paris, Graduate School of Engineering, Paris, F-75015, France}

\author{Etienne Mangaud}
 \affiliation{MSME, Universit\'e Gustave Eiffel, UPEC, CNRS, F-77454 Marne-La-Vall\'ee, France}

\author{Alex W. Chin}
\affiliation{Institut des Nanosciences de Paris, Sorbonne Université, CNRS, F-75005 Paris, France}

\author{Mich\`ele Desouter-Lecomte}
\email{michele.desouter-lecomte@universite-paris-saclay.fr}
\affiliation{Institut de Chimie Physique, Universit\'e Paris-Saclay-CNRS, UMR8000, F-91400 Orsay, France} 


\date{\today}

\begin{abstract}
In non-perturbative non-Markovian open quantum systems, reaching either low temperatures with the hierarchical equations of motion (HEOM) or high temperatures with \textcolor{black}{the Thermalized Time Evolving Density Operator with Orthogonal Polynomials (T-TEDOPA) formalism in Hilbert space remains challenging.} We compare different manners of modeling the environment. Sampling the Fourier transform of the bath correlation function, also called temperature dependent spectral density, proves to be very effective. T-TEDOPA (Tamascelli \textit{et} \textit{al}. Phys. Rev. Lett. 123, 090402 (2019)) uses a linear chain of oscillators with positive and negative frequencies while HEOM is based on the complex poles of an optimized rational decomposition of the temperature dependent spectral density (Xu \textit{et} \textit{al}. Phys. Rev. Lett. 129, 230601 (2022)). Resorting to the poles of the temperature independent spectral density and of the Bose function separately is an alternative when the problem due to the huge number of the Bose poles at low temperature is circumvented. Two examples illustrate the effectiveness of the HEOM and T-TEDOPA approaches: a benchmark pure dephasing case and a two-bath model simulating dynamics of excited electronic states coupled through a conical intersection. We show the efficiency of T-TEDOPA to simulate dynamics at a finite temperature by using either continuous spectral densities or only all the intramolecular oscillators of a linear vibronic model calibrated from \textit{ab} \textit{initio} data of a phenylene ethynylene dimer. 
\end{abstract}


\maketitle 

\section{Introduction}
\textcolor{black}{Increasing temperature in the multidimensional wave function formalism or decreasing it in an open quantum system described by a reduced density matrix is a challenge giving rise to many methodological developments. In the Hilbert space, the wave function is $\textit{a}$ $\textit{priori}$ at zero Kelvin. In the Liouville space, the reduced density matrix describes the main degrees of freedom providing the observations of interest. It is obtained by tracing all the surrounding degrees of freedom. The environment is present in the master equation only via statistical temperature dependent correlation functions. This is particularly well adapted to treat thermal bosonic baths at room or higher temperatures. In this work, we focus on two methods: the Hierarchical Equations of Motion (HEOM) \cite{Kubo1989,Tanimura2006,Tanimura2020, Yan2007} in Liouville space and the discrete chain mapping T-TEDOPA algorithm (Thermalized Time Evolving Density Operator with Orthogonal Polynomials) \cite{Tamascelli2019,Tamascelli2020,Plenio2024} in Hilbert space. We survey some recent methods to optimize the computation at low temperature with HEOM and we compare with the T-TEDOPA approach to enlighten its performance at high temperature.}     

\textcolor{black}{HEOM is a reference method for non-perturbative non-Markovian dynamics \cite{Ishizaki_Fleming2009}, i.e., when the system is strongly coupled to the surrounding and when the characteristic lifetime of the dynamics in both spaces is of the same order of magnitude. The HEOM are exact for baths with Gaussian statistics (harmonic baths) \cite{Kubo1989,Ishizaki2005,Tanimura2006,Tanimura2020, Yan2007}. The method is essentially based on a parametrization of the bath two-time correlation function in terms of a sum of complex decaying exponential functions. Each term determines an artificial mode that can absorb or emit energy. The algorithm consists in an ensemble of coupled equations among auxiliary density operators (ADOs) having the same dimension as the system reduced matrix. Each ADO is associated with a given occupation number in the effective modes representing the thermal reservoir. The ADOs are coupled when they may exchange one quantum to increase or decrease the excitation. The ADO hierarchy is infinite in principle but truncated in practice by limiting the maximum occupation number. Reviewing $\textit{in}$ $\textit{extenso}$ the wide range of algorithms aimed at enhancing HEOM performance is outside the scope of this work. For instance, algorithms concern the filtering of the ADOs to reduce their number \cite{Ishizaki2005,Yan2009} or the consideration of an initial system-bath correlation when the initial total density matrix cannot be factorized into a system matrix and a Boltzmann equilibrium bath density \cite{Tanimura2014,Shi2015}. Here, we tackle the problem of the low temperature regime when HEOM would require a huge number of modes   \cite{Ishizaki2005,Duan2017,Lambert2019,Fay2022,Xu2022}.}

\textcolor{black}{
A wide variety of possibilities exists to choose artificial modes in the time or frequency domains by acting directly on the correlation function \cite{Wu2015,Nakamura2018,Lambert2019,Kleinekathofer2019,Ikeda2020,Plenio2020,Yan2022,Lambert2023,ThossBorrelli2024} or on its Fourier transform. The latter is the product of the temperature independent spectral density giving the system-bath coupling strength as a function of the frequency and the temperature dependent Bose function \cite{Breuer2002,Mukamel1995}. The product may be called the temperature dependent spectral density. We compare three types of artificial decay modes for HEOM. First, the correlation function is parametrized in terms of the poles in the complex plane of both the spectral density and the Bose function. Second, following a more recent procedure proposed by M. Xu $\textit{et}$ $\textit{al}.$ we use the poles of the temperature dependent spectral density directly. This method is called FP-HEOM (Free-Pole HEOM) \cite{Xu2022,Krug2023}. It is based on the barycenter representation \cite{Nakatsukasa2018} of this temperature dependent spectral density approximated by a rational fraction. This extension of HEOM  has also been successfully applied to fermionic baths \cite{Dan2023}. Finally, we consider a sampling of the spectral density on the real frequency axis with undamped oscillators, }  
   
\textcolor{black}{In the Hilbert space, finite temperature was introduced in various ways, for instance by sampling initial conditions over a Wigner distribution \cite{Burghardt2020} or using stochastic thermal wave function in the MCTDH (Multi Configuration Time Dependent Hartree)  or ML-MCTDH (Multi-Layer MCTDH) simulations \cite{Tremblay2022}. Thermo-Field dynamics has also been coupled with MCTDH \cite{Saalfrank2014,Borrelli2021}, Davidov ansatz \cite{Zhao2017,Zhao2023} or propagation with the matrix product state (MPS) representation   \cite{deVega2015,Gelin2023}. Recently the T-TEDOPA algorithm \cite{Tamascelli2019,Tamascelli2020,Dunnett2021,Plenio2024} was found to succeed in simulating a thermal environment with wave functions at 0 K propagated with MPS \cite{Chin2016}. The thermal environment is mapped on a linear chain of oscillators \cite{ Dunnett2021,Riva2023}. The temperature dependent spectral density spanning the axis of negative and positive frequencies is then the main tool to determine the different couplings. Using of the temperature dependent spectral density is a common point between FP-HEOM and T-TEDOPA. However, FP-HEOM uses poles in the complex plane and a "star model" in which each mode is coupled to the system whereas T-TEDOPA is based on a "chain model". In this case, the system is interacting with a single mode that is the first mode of a chain of coupled neighboring oscillators \cite{Chin2010,Christ2009,Burghardt2010,Pereverzev2009,Eisfeld2012,Plenio2019}.
We do not consider here the reaction coordinate mapping method where one or two main modes are included in the active system \cite{Chenel2014,Mangaud2015,Nazir2014,Nazir2016,Gelin2016}. Both HEOM and T-TEDOPA methods benefit from implementing with the MPS representation, which reduces the computational resources and is a good candidate to handle the curse of dimensionality \cite{Shi2018,Shi2020,Shi2021,Dolgov2021,Thoss2022,Ke2023,Mangaud2023}.}

The paper is organized as follows. In Sec. \ref{sec:modes}, \textcolor{black}{we define the main tools in open quantum dynamics: the bath correlation function, the spectral density in frequency domain and the temperature dependent spectral density. We give an introduction to HEOM method in the conventional implementation and in the MPS representation. We then define three kinds of artificial decay modes involved in the expansion of the bath correlation function in HEOM.} Section \ref{sec:TWPmodes} presents the sampling of the temperature dependent spectral density for T-TEDOPA. \textcolor{black}{Numerical details for HEOM or T-TEDOPA propagation are gathered in Sec.\ref{sec:nummethod}.} The simulations are compared in Sec.\ref{sec:PD} through a benchmark case involving a qubit in a pure dephasing environment for which an analytical solution exists \cite{Breuer2002}. Section \ref{sec:twobath} illustrates a two-bath case with strong system-bath coupling and non-Markovian dynamics. One bath is coupled diagonally making fluctuate the energies of the system while the other one is coupled off-diagonally altering the inter-state coupling. This situation is typical of a conical intersection between two excited electronic states. The model is calibrated here from an $\textit{ab}$ $\textit{initio}$ investigation of a phenylene ethynylene dimer (1,3-bis(phenylethynyl)benzene) by B. Lasorne $\textit{et}$ $\textit{al.}$ \cite{Ho2019,Jaouadi2022}.  We compare dynamics at a finite temperature in condensed phase with continuous spectral densities or using only the intramolecular vibrators of the $\textit{ab}$ $\textit{initio}$ linear vibronic coupling (LVC) model.  Finally section \ref{sec:conclu} gives some concluding remarks.

\section{Artificial decay modes in HEOM}
\label{sec:modes}
\textcolor{black}{We set $\hbar = 1$ and we adopt mass weighted coordinates throughout the paper.} The generic Hamiltonian of a complex system is usually split into three parts 
\begin{equation}
H=H_S+H_{SB}+H_B.    
\end{equation}
$H_S$ describes the central system containing the main active degrees of freedom. \textcolor{black}{The environmental Hamiltonian
\begin{equation}
{{H}_{B}}=\frac{1 }{2}\sum\nolimits_{\alpha }^{{{N}_{bath}}}{\sum\nolimits_{j}^{{{N}_{\alpha }}}{\omega _{\alpha j}^{2}}q_{\alpha j}^{2}}
\label{eq:HB}
\end{equation}
is a collection of harmonic oscillators possibly separated in different $N_{bath}$ baths. The system-bath coupling is
\begin{equation}
{{H}_{SB}}=\sum\nolimits_{\alpha =1}^{{{N}_{bath}}}{{{S}_{\alpha }}}{{B}_{\alpha }}
\label{eq:HSB}
\end{equation}
where ${{S}_{\alpha }}$ is an operator in the system space
\begin{equation}
{B}_{\alpha }=\sum\nolimits_{j}^{{{N}_{\alpha }}}c_{\alpha j}{q}_{\alpha j}   
\end{equation}
 is a collective bath coordinate  built with the system-bath couplings $c_{\alpha j}$. The latter are the vibronic couplings when the partition concerns an electronic system and the vibrational motions.  For sake of simplicity, we give the following relations by considering a single bath, thus dropping the $\alpha$ index. The generalization does not pose any major difficulties. The main tool of a non-Markovian master equation simulating an open quantum system with Gaussian statistics is the two-time correlation function of the thermal bath $C(t)={{\left\langle B(t)B(0) \right\rangle }_{eq}}$ where $B(t)$ is the Heisenberg representation of the operator and ${{\left\langle \centerdot  \right\rangle }_{eq}}$ denotes the average over a Boltzmann distribution at temperature $T$.} The correlation function is related to the spectral density by the following relation:
\begin{equation}
 C(t)=\frac{1}{\pi }\int_{-\infty }^{\infty }\mathrm{d} \omega J(\omega )(1+n_{\beta}(\omega)){{e}^{-i\omega t}}
\label{eq:Cdet}
\end{equation}
where 
\begin{equation}
 {{n}_{\beta }}(\omega )=1/(e^{\beta \omega}-1)
\label{eq:Bose}
\end{equation} 
 is the Bose function with $\beta =1/{{k}_{B}}T$ and $k_B$ is the Boltzmann constant. \textcolor{black}{$J(\omega )$ is the temperature independent spectral density specifying the coupling to the environment at each frequency. By assuming a continuous distribution in frequency, it reads 
\begin{equation} 
 J(\omega )=\frac{\pi }{2}\int_{-\infty }^{\infty }{\frac{{{c}^{2}}(\omega ')}{\omega '}}\delta (\omega -\omega ')\mathrm{d}\omega '. 
 \label{eq:J(w)}    
 \end{equation} }
This expression satisfies the relation $J(-\omega)=-J(\omega)$. The power spectrum of $C(t)$
 \begin{equation}
 {{J}_{\beta }}\left( \omega  \right) = J(\omega )(1+n_{\beta}(\omega))
 \label{eq:Jbeta}    
 \end{equation}
is the temperature dependent spectral density. The central point in HEOM is the expansion of $C(t)$ and of its complex conjugate $\bar{C}(t)$ as a sum of contributions. The more popular expression is a sum of decaying exponential functions
\begin{equation}
C(t)=\sum\nolimits_{k=1}^K{{{\alpha }_{k}}}{{e}^{i{{\gamma }_{k}}t}}
\label{eq:Cdetexpan}
\end{equation}
with complex coefficients $\alpha_k$ and $\gamma_{k} = \Omega_k+i\Gamma_k$. Each term corresponds to a bath artificial decay mode with positive or negative frequency $\Omega_k$ and positive decay rate $\Gamma_k$. Different expansions have also been proposed \cite{Yan2010,Wu2015,Nakamura2018,Lambert2019,Kleinekathofer2019,Ikeda2020,Yan2022}. 

\textcolor{black}{The choice of the modes is crucial since the computational cost may dramatically increase with the total number $K$ as we discuss below. We first summarize the generic structure of the HEOM coupled equations used in a conventional implementation or in the MPS format \cite{Shi2018,Shi2020,Shi2021,Borrelli2021,Ke2023,Mangaud2023} to show how the complexity increases with $K$ and with the strength of the system-bath coupling. We then survey three possible choices of decay modes involving poles of $J(\omega)$ or  $J_{\beta}(\omega)$ to get around the dimensionality problem by some efficient procedures. The particular operational equations are given in the Supplementary Material. }

\begin{figure}
    \centering
    \includegraphics [width =1\columnwidth, height = 6 cm]{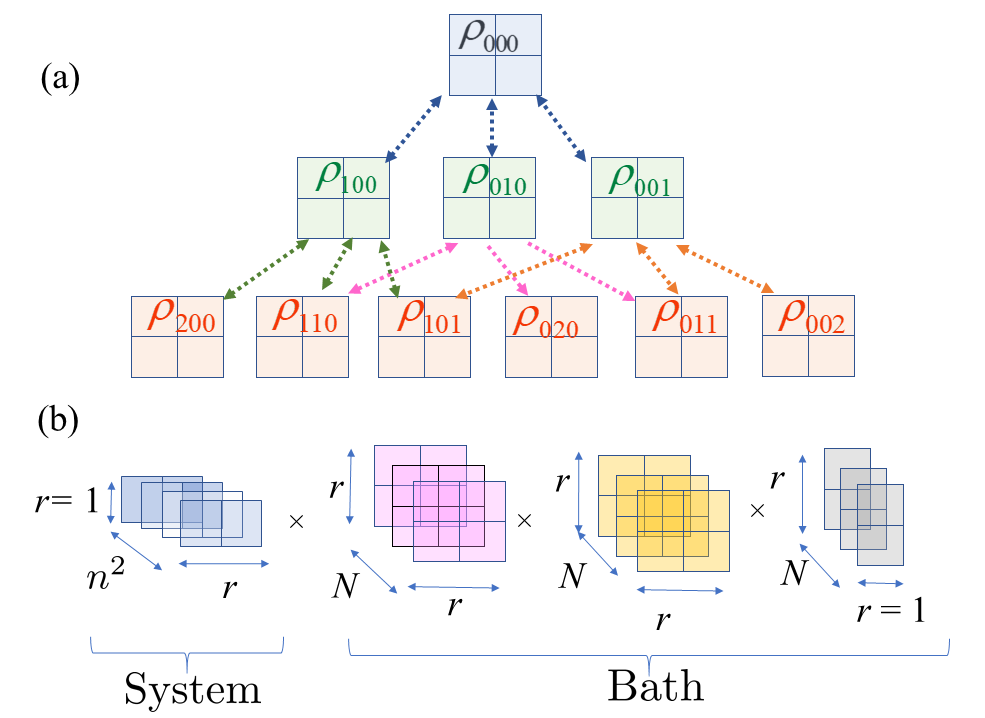}
    \caption{(a) Schematic representation of a HEOM hierarchy for an open system with dimension $n=2$, three artificial modes $K=3$ and a truncation at level $L=2$ corresponding to a Fock space $N=3$ for each mode. (b) Representation of the supervector containing the system matrix and the ADOs in the MPS format by assuming a bond $r=2$.  }
    \label{fig:fig1}
\end{figure}

\textcolor{black}{The initial total density operator
\begin{equation}
\rho_{tot}(t=0) = \rho_S(t=0) \otimes \rho_{B,eq}
\label{rhoini}    
\end{equation}
is assumed to be factorized into the product of an arbitrary system density operator $\rho_S$ and a thermal equilibrium bath density operator $\rho_{B,eq}=e^{-\beta H_B}/Tr_B\left[e^{-\beta H_B}\right]$. Correlated initial states may also be considered \cite{Shi2015,Tanimura2014}. The interaction with the bath is treated by a time local system of coupled equations among ADOs organized in a hierarchical structure. Each ADO has the dimension of the system density matrix and is labelled by a global index vector $\mathbf{m} = (m_1,..,m_k,...,m_K)$ giving the number of occupation in each artificial mode. The equations take the form:
\begin{align}
  & {{{\dot{\rho }}}_{\mathbf{m}}}(t)={{L}_{S}}{{\rho }_{\mathbf{m}}}(t)+i\sum\limits_{k=1}^{K}{{{m}_{k}}{{\gamma }_{k}}{{\rho }_{\mathbf{m}}}}(t) \nonumber \\ 
 & -i\left[ S,\sum\limits_{k=1}^{K}{{{\rho }_{\mathbf{m}_{k}^{+}}}(t)} \right] \nonumber \\ 
 & -i\sum\limits_{k=1}^{K}{{{m}_{k}}\left( {{\alpha }_{k}}S{{\rho }_{\mathbf{m}_{k}^{-}}}(t)-{{{\tilde{\alpha }}}_{k}}{{\rho }_{\mathbf{m}_{k}^{-}}}(t)S \right)} 
\label{eq:eqcouHEOM}
\end{align}
where $\mathbf{m}_{k}^{\pm }=\{{{m}_{1,}}...,{{m}_{k}}\pm 1,..{{m}_{M}}\}$ are the index of ADOs for which the occupation number has changed by one unit. The hierarchy is thus structured in layers characterized by a given total occupation number. Each level interacts only with the two neighboring layers. The hierarchy is schematized in Fig.\ref{fig:fig1}(a) for a simple example with a two-dimensional system matrix $n=2$, three artificial modes $K=3$ and a truncation at level $L=2$ where $L$ designates the maximum excitation in each decay mode. The equations contain the parameters of the expansion of $C(t)$ [Eq.(\ref{eq:Cdetexpan})]. The definition of the $\tilde{\alpha}_{k}$ for the different choices of the artificial modes are given in the Supplemental Material.  When the hierarchy is truncated at a given level $L$, the number of ADOs estimated without any filtering procedure \cite{Ishizaki2005,Yan2009} may become dramatically large. The maximum level $L$ is intimately linked to the strength of the system-bath coupling that favours extensive energy exchange between the system and the $K$ artificial modes. For a $n-$dimensional system, the standard implementation without any filtering requires the storage of  
\begin{equation}
N_{HEOM}^{st}=n^2  ( L+K)!/L!K!
\label{eq:NHEOMst}    
\end{equation}
 complex matrix elements. 
}

\textcolor{black}{A promising alternative is the HEOM implementation with the MPS format \cite{Shi2018,Shi2020,Shi2021,Borrelli2021,Ke2023,Mangaud2023} for which we now summarize the main expressions. Each matrix of dimension $n \times n$ is transformed into a $n^2$ vector. The full multi-dimensional array $\rho_\mathbf{m}$ becomes a super vector $\tilde{\rho}_\mathbf{m}$ on which act the superoperators corresponding to the different terms of the master equation (\ref{eq:eqcouHEOM}) that is written now
${{\dot{\tilde{\rho }}}_{\mathbf{m}}}(t)=\mathcal{L}{{\tilde{\rho }}_{\mathbf{m}}}(t)$ with
\begin{equation}
\mathcal{L} = \mathcal{L}_S + \sum_{k=1}^{K} \left( \mathcal{L}_{k} + \mathcal{L}_{k}^+ + \mathcal{L}_{k}^- \right) 
\label{eq:lioutt}
\end{equation} 
where $\mathcal{L}_{k}$ is the damping term involving the complex $\gamma_k$ rates, $\mathcal{L}_{k}^+$ and $\mathcal{L}_{k}^-$ concern the terms involving the upper hierarchy level or the lower one respectively. We consider a generic case with matrices $n \times n$ and $K$ artificial modes for which the maximum occupation number is $n_{max}$ leading to a local Fock space of dimension $N = n_{max}+1= L+1$. $\tilde{\rho}_\mathbf{m}$ is built as a train of $K+1$ cores that are three dimensional arrays. The dimension of a core $A_k$ is $r_k \times N \times r_{k'}$ where $r_k$ and $r_{k'}$ are the $k$th and $k'$th bonds that must be carefully calibrated. We choose ${{r}_{k}}=r$  for all $k\in \left[\!\left[ 1,K \right]\!\right]$. The first ($A_0$) and the last ($A_K$) cores have different dimensions $1 \times n^2 \times r$ and $r \times N \times 1$ respectively. The tensor train format of the example given in Fig.\ref{fig:fig1}(a) is illustrated in Fig.\ref{fig:fig1}(b) by assuming that the bond is $r=2$. A given element $J = ij$ of the supervector corresponding to a given matrix with $\mathbf{m} = (m_1,..,m_k,...,m_K)$ is expressed as:
\begin{align}
  & \tilde{\rho }_{{{m}_{1}},..,{{m}_{k}},..,{{m}_{K}}}^{J }=\sum\limits_{{{j}_{1}}}{\sum\limits_{{{j}_{2}}}{..\sum\limits_{{{j}_{k}}}{..\sum\limits_{{{j}_{K}}}{{{A}_{0}}(J ,{{j}_{1}})}}}}  \nonumber \\ 
 & \times {{A}_{1}}({{j}_{1}},{{m}_{1}},{{j}_{2}})...{{A}_{k}}({{j}_{k}},{{m}_{k}},{{j}_{k+1}})...\nonumber  \\ 
 & \times {{A}_{K}}({{j}_{K}},{{m}_{K}}). 
\end{align}
To explicit the different Liouvillian operators, we define a $n \times n$ identity matrix $\mathbf{I}_{n}$, $K$ $N \times N$ identity matrices $\mathbf{I}_{N_j}$ and a supervector $\mathbf{I}_{n^2}$. We also introduce three $N \times N$ matrices $\mathbf{M}_{k}$, $\mathbf{M}_{k}^{'}$ and  $\mathbf{M}_{k}^{''}$ defined after the different operators. These are written with the Kronecker product symbol $\otimes$ as follows:
\begin{equation}
{\mathcal{L}_{S}}=-i\left( {{H}_{S}}\otimes {{\mathbf{I}}_{n}}-{{\mathbf{I}}_{n}}\otimes {{H}_{S}} \right)\otimes \prod\limits_{j=1}^{K}{{\mathbf{I}_{{{N}_{j}}}}}    
\end{equation}
\begin{equation}
{\mathcal{{L}}_{k}}=i{{\gamma }_{k}}{\mathbf{I}_{{{n}^{2}}}}\otimes \prod\limits_{j=1}^{k-1}{{\mathbf{I}_{{{N}_{j}}}}}\otimes {\mathbf{M}_{k}}\otimes \prod\limits_{j=k+1}^{K}{{\mathbf{I}_{{{N}_{j}}}}}   
\end{equation}
with a diagonal matrix ${{\left( {{\mathbf{M}}_{k}} \right)}_{l,l}}=l-1$,
\begin{align}
  &\mathcal{L}_{_{k}}^{+}=-i\left( S\otimes {{\mathbf{I}}_{n}}-{{\mathbf{I}}_{n}}\otimes S \right) \nonumber \\ 
 & \otimes \prod\limits_{j=1}^{k-1}{{{I}_{{{N}_{j}}}}}\otimes \mathbf{M}_{k}^{'}\otimes \prod\limits_{j=k+1}^{K}{{{I}_{{{N}_{j}}}}}  
\end{align}
with only the first upper diagonal ${{\left( \mathbf{M}_{k}^{'} \right)}_{l,l+1}}=1$,
\begin{align}  \label{eq:Lmoins}
  & \mathcal{L}_{_{k}}^{-}=-i\left( {{\alpha }_{k}}S\otimes {{\mathbf{I}}_{n}}-\Tilde{\alpha}_{k}{{\mathbf{I}}_{n}}\otimes S \right)  \nonumber \\ 
 & \otimes \prod\limits_{j=1}^{k-1}{{{I}_{{{N}_{j}}}}}\otimes \mathbf{M}_{k}^{''}\otimes \prod\limits_{j=k+1}^{K}{{{I}_{{{N}_{j}}}}}  
\end{align}
with only the first lower diagonal ${{\left( \mathbf{M}_{k}^{''} \right)}_{l,l-1}}=l$. When the cores have the same bond dimension $r$ as assumed here, the MPS format involves
\begin{equation}
N_{HEOM}^{MPS}=r^2  N  (K-1) +r(n^2+N)
\label{eq:NHEOMMPS}   
\end{equation}
complex elements. }

\textcolor{black}{We now summarize three strategies to make the number $K$ of artificial modes in HEOM applications as small as possible by acting on the spectral density.}

\subsection{Modes from the poles of $J(\omega)$}
\label{sec:TMmodes}
\textcolor{black}{This strategy consists in considering separately the spectral density $J(\omega)$ and the Bose function of the $C(t)$ power spectrum [Eq.(\ref{eq:Cdet})]. The poles in the complex plane allow the computation of the integral (\ref{eq:Cdet}) by the residue theorem. Fitting $J(\omega)$ by some functions provides analytical expressions for the $\alpha_k$ and $\gamma_k$ coefficients. The antisymmetry of $J(\omega)$ and the fact that it is real on the frequency axis leads to some structure in the distribution of the poles, which occur in conjugated pairs. The poles of the Bose function, also called Matsubara frequencies, are also known exactly and located on the imaginary axis at $\omega = 0$.}

\begin{figure}
    \centering
    \includegraphics [width =1\columnwidth, height = 6 cm]{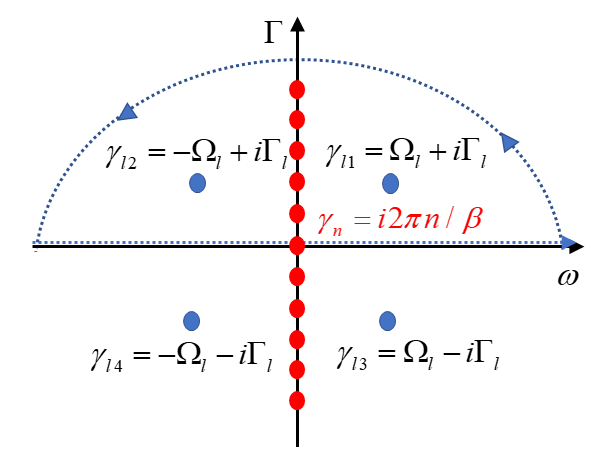}
    \caption{Schematic representation of the four poles of a Tannor-Meier Lorentzian function [Eq.(\ref{eq:JTM})] in the complex plane. The poles of the Bose function on the imaginary axis are red circles. The contour for the calculation of the correlation function [Eq.(\ref{eq:Cdet})] by the residue theorem is in dotted line.  }
    \label{fig:fig2}
\end{figure}

\textcolor{black}{An early method reviewed in references \cite{Shi2014,Eisfeld2014} consists in fitting $J(\omega)$ by a Drude-Lorentz function \cite{Kubo1989,Tanimura2006} or by a combination of $N_l$ Tannor-Meier (TM) Lorentzian functions \cite{Meier1999,Pomyalov2010}
\begin{equation}
 J^{TM}  \approx\sum\limits_{l=1}^{{{N}_{l}}}{\frac{{{p}_{l}}\omega }{\left[ {{\left( \omega +{{\Omega }_{l}} \right)}^{2}}+\Gamma _{l}^{2} \right]\left[ {{\left( \omega -{{\Omega }_{l}} \right)}^{2}}+\Gamma _{l}^{2} \right]}}.
\label{eq:JTM}
\end{equation}
These non-symmetrical functions are well adapted to describe a linear behavior at low energies of an Ohmic spectral density. They are also called Ohmic-Lorentzian functions. Moreover they effectively fit highly structured spectral densities obtained when some environmental modes are strongly coupled to the system leading to sharp peaks. The extension to super-Ohmic Lorentzian functions behaving as $\omega^3$ at low frequencies is given in Ref.\cite{MangaudMeier2017}. Each Ohmic-Lorentzian function labelled with subscript $l$ has four single poles as schematized in Fig.\ref{fig:fig2}. The contour for the residue theorem lies along the whole real $\omega$ axis and is closed in the upper half-plane so that only two poles provide two artificial decay modes with opposite frequencies} 
\begin{align}  \label{eq:gammaTM}
 {{\gamma^{TM}_{l1}=\Omega }_{l}} +i{{\Gamma }_{l}} \nonumber \\ 
 {{\gamma^{TM}_{l2}=-\Omega }_{l}}+i{{\Gamma }_{l}}.
\end{align} 
The corresponding $\alpha_k$ complex coefficients of Eq.(\ref{eq:Cdetexpan}) are then:
\begin{align}  \label{eq:alphaTM}
  & {{\alpha }^{TM}_{l1}}=\frac{{{p}_{l}}}{8{{\Omega }_{l}}{{\Gamma }_{l}}}\left[ \coth \left( \frac{\beta }{2}({{\Omega }_{l}}+i{{\Gamma }_{l}}) \right)-1 \right] \nonumber \\  
 & {{\alpha }^{TM}_{l2}}=\frac{{{p}_{l}}}{8{{\Omega }_{l}}{{\Gamma }_{l}}}\left[ \coth \left( \frac{\beta }{2}({{\Omega }_{l}}-i{{\Gamma }_{l}}) \right)+1 \right]. 
\end{align}

The low temperature problem comes from the infinite number of Matsubara frequencies that are the poles of the Bose function on the imaginary axis 
\begin{equation}
 \gamma^{TM}_{n} = i \nu_{n}
 \label{eq:gamMatsu}
\end{equation}
where $\nu_{n}=2\pi n/\beta$ with $n$ an integer. The corresponding $\alpha^{TM}_k$ coefficients for the Matsubara frequencies on the positive imaginary axis are 
\begin{equation}
 \alpha^{TM}_{n}=2iJ^{TM}(i\nu_{n})/\beta=2iJ^{TM}(\gamma_n^{TM})/\beta.
 \label{eq:alphaMatsu}
\end{equation}
\textcolor{black}{The correlation function [Eq.(\ref{eq:Cdetexpan})] can be rewritten as the sum of the contributions from the two kinds of poles
\begin{equation}
{{C}^{TM}}(t) \approx C_{Lor}^{{}}(t)+C_{Matsu}^{{}}(t)    
\end{equation}
with
\begin{equation}
{{C}_{Lor}}(t)=\sum\limits_{j=1,2}{\sum\limits_{l=1}^{{{N}_{l}}}{\alpha _{lj}^{TM}}}{{e}^{i\gamma _{lj}^{TM}t}}    
\end{equation}
and
\begin{equation}
C_{Matsu}(t)=\sum\nolimits_{n=1}^{{{M}_{a}}}{{{\alpha }^{TM}_{n}}}{{e}^{-{{\nu }_{n}}t}}
\label{eq:CMdet}    
\end{equation}
where $M_a$ is infinite in principle.
Fortunately, the infinite series may be drastically truncated at high temperature by retaining only very few $M_a$ Matsubara terms. The total number of artificial modes in this TM parametrization is then
\begin{equation}
K^{TM}=2N_l+M_a.
\label{eq:MTM}    
\end{equation}
This explains the efficient utilization of HEOM at room temperature. However, decreasing the temperature readily requires hundreds of terms to get convergence. To circumvent this difficulty another expansion of the Bose function near zero temperature has been proposed in terms of Fano spectrum decomposition \cite{Cui2019}.  Recently, N. Lambert $\textit{et}$ $\textit{al}.$\cite{Lambert2019} suggested to fit the Matsubara contribution of the correlation function [Eq.(\ref{eq:Cdetexpan})]. This function $C_{Matsu}(t)$ [Eq.(\ref{eq:CMdet})] is real and negative and may be fitted by a sum of a small number of real exponential functions:
\begin{equation}
C_{Matsu}^{fit}(t)\approx\sum\limits_{k=1}^{M_{fit}}a_{k}\times e^{(-b_{k}t)}.
\label{eq:CMdetfit}
\end{equation}
This approach will be denoted in our study as TM\&FIT (Tannor-Meier with Fitted Matsubara modes). The total number of artificial modes is then
\begin{equation}
K^{TM\&FIT}=2N_l+M_{fit}
\label{eq:MTMFIT}    
\end{equation}
with $M_{fit}<<M_a$.}

\subsection{Modes from the poles of $J_{\beta}(\omega$)}
\label{sec:FPmodes}
\begin{figure}
    \centering
    \includegraphics [width =1.\columnwidth, height = 4 cm]{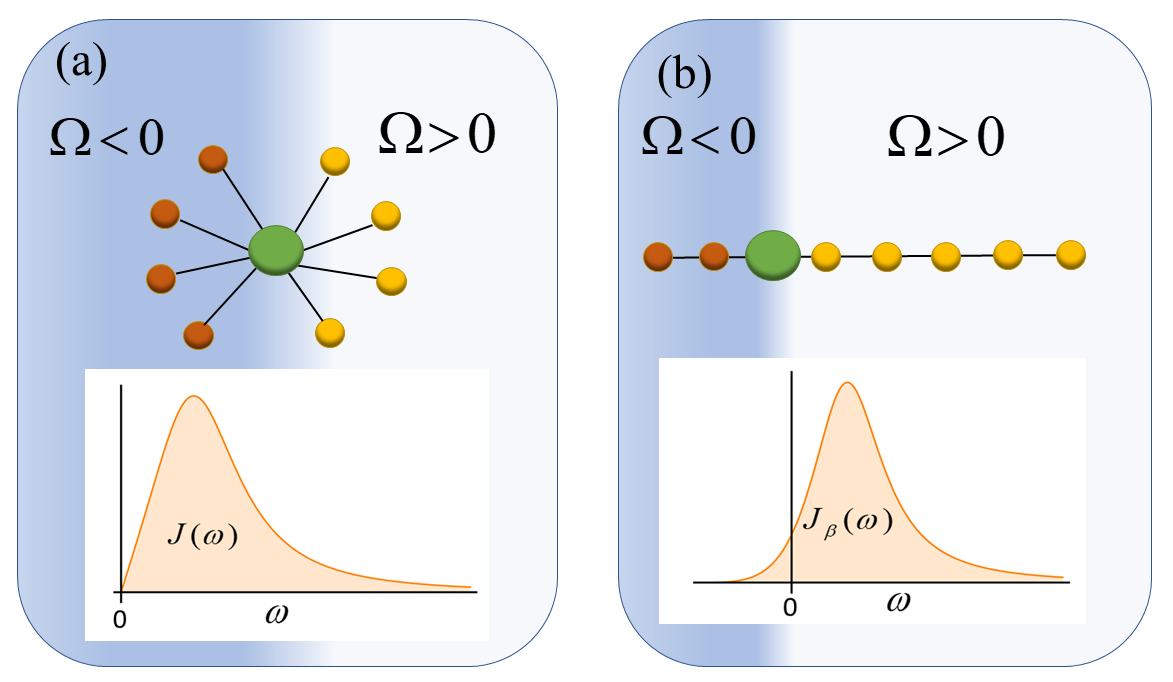}
    \caption{ (a) Discrete modes in the HEOM or T-TEDOPA simulations. The star model of HEOM is built with modes sampling $J(\omega)$ with $\omega > 0 $ and the bath correlation function [Eq.(\ref{eq:Cdetexpan})] involves an equal number of positive and negative frequencies with weights given in Eq.(\ref{eq:alphaMD}). (b) The discrete modes of T-TEDOPA samples $ J_{\beta}(\omega)$ with a minimum cutoff $\omega_{min}$ leading to a dissymetric number of modes with positive or negative frequencies. A unitary transformation replaces the star model to a chain model of coupled oscillators. }
    \label{fig:fig3}
\end{figure}

\textcolor{black}{Recently, a very interesting procedure has been proposed by M. Xu $\textit{et}$ $\textit{al.}$ \cite{Xu2022} by using only the poles the temperature dependent ${{J}_{\beta }}\left( \omega  \right)$ to compute the Fourier transform [Eq.(\ref{eq:Cdet})]. The density is approximated by a the barycentric representation of a rational function
\begin{equation}
 {J}^{FP}_{\beta }(\omega )\approx\sum\limits_{n=1}^{{{m}_{p}}}{\frac{{{w}_{n}}{{J}_{\beta }}({{\Omega }_{n}})}{\omega -{{\Omega }_{n}}}}/\sum\limits_{n=1}^{{{m}_{p}}}{\frac{{{w}_{n}}}{\omega -{{\Omega }_{n}}}}
\label{eq.AAA}    
\end{equation}
where $w_n$ are real or complex weights. ${J}_{\beta }(\omega )$ is not antisymmetric as is ${J}(\omega )$. The pole distribution loses its symmetry for positive and negative frequencies. However, as ${J}_{\beta }(\omega )$ is real on the real axis, the poles still occur in conjugated pairs. Starting from a large sampling of ${J}_{\beta }(\omega )$, the number of terms in Eq.(\ref{eq.AAA}) is optimized iteratively from $m_p=1$ to a value according to a tolerance for the error of the fit. The optimization uses the Adaptive Antoulas-Anderson (AAA) algorithm \cite{Nakatsukasa2018} and the  BARYRAT algorithm \cite{Hofreither2021} with Python packages  \cite{AAA,baryrat}. This rational barycentric interpolation provides a small number of accurate poles $P_k = \Omega_k^{FP}+i\Gamma_k^{FP}$ of ${{J}_{\beta }}\left( \omega  \right)$ that are determined as roots of a polynomial. The expansion of the correlation function is obtained by contour integration [Eq.(\ref{eq:Cdet})] using the optimized $M_{p}=m_p/2$ poles $P_k$ in the upper half-plane. Their frequencies $\Omega_k^{FP}$ are real-valued and the rates $\Gamma_k^{FP}$ are positive. The minimum number of poles is fixed by computing the correlation function [Eq.(\ref{eq:Cdetexpan})] until a characteristic time of the dynamics. The number of artificial modes involved in FP-HEOM is
\begin{equation}
K^{FP}=2M_p
\label{eq:MmodesFP}   
\end{equation}
including for each mode of frequency $\Omega_k^{FP}$ a mode with frequency $-\Omega_k^{FP}$. The decay rates of a pair are
 \begin{align}
 &{{\gamma }_{Pk1}}=\Omega_k^{FP}+i\Gamma_k^{FP} \nonumber \\
 &{{\gamma }_{Pk2}}=-\Omega_k^{FP}+i\Gamma_k^{FP}
 \label{eq:gammaFP}    
 \end{align}
and the corresponding ${{\alpha }_{Pk}}$ values are 
 \begin{align}
  & {{\alpha }_{k1}^{FP}} = 2i Res[P_k]  \nonumber \\
  & {\alpha }_{k2}^{FP} = {\bar{\alpha }}_{k1}^{FP}
  \label{eq:alphaFP}
 \end{align}
 where $Res[P_k]$ is the residue at each selected pole. Then, Eq.(\ref{eq:Cdetexpan}) writes
 \begin{equation}
C(t)\approx \sum\limits_{j=1,2}{\sum\limits_{k=1}^{{{M}_{p}}}{\alpha _{kj}^{FP}}}{{e}^{i\gamma _{kj}^{FP}t}}.     
 \end{equation}
 }
 \textcolor{black}{
\subsection{Discretization of $J(\omega)$}
\label{sec:Dmodes}
Discretization of the continuous spectral density $J(\omega)$ to map the total system onto a spin-Boson model with discrete undamped modes is a common practice in multidimensional MCTDH wave function simulations but is used also in HEOM applications \cite{Shi2014,Shi2018,Shi2021}, with the generalized quantum master equation \cite{Geva2019} or in semi-classical simulation \cite{Geva2020}. The first clear distinction is based on the mapping of $J(\omega )$, for instance in MCTDH applications \cite{Thoss2001,Burghardt2012,Burghardt2020,Lasorne2024} or sparse grid method \cite{Lauvergnat2023} \textit{versus} $J_{\beta}(\omega )$ in the T-TEDOPA method discussed in Sec.\ref{sec:TWPmodes}. The second main difference concerns the resulting discrete model that is a star model when all the modes are directly coupled to the system or a chain model when they form a hierarchical chain of coupled neighboring modes as schematized in Fig.\ref{fig:fig3}. In this section devoted to HEOM simulations denoted by D-HEOM, we consider the mapping of $J(\omega )$ on a star model. The discretized spectral density is then written:
\begin{equation}
{{J}^{D}}(\omega )=\frac{\pi }{2}\sum\limits_{m=1}^{M}{\frac{c_{m}^{D2}}{{{\omega }_{m}^D}}\delta (\omega -{{\omega }_{m}^D})}.
\label{eq:Jdiscret}
\end{equation}
}
There are obviously different discretization strategies to obtain the $c_m^D$ and $\omega_m^D$, see for instance references \cite{deVega-Wolf2015,Makri2017,Christ2009,Pereverzev2009,Burghardt2010}. \textcolor{black}{The objective is to work with as few modes as possible and to obtain a correct representation of the correlation function at least until a relevant time cutoff \cite{Plenio2024}. We have tried the technique early used in MCTDH \cite{Thoss2001} with regular spacing and a coupling strength weighted by the local state density but the convergence of $C(t)$ was slower than with a non-uniform spacing discretization of $J(\omega )$ suggested in Ref. \cite{Makri2017}.} This procedure ensures that all modes contribute to the same fraction of the reorganization energy
\begin{equation}
\lambda =\frac{1}{\pi }\int_{0}^{\infty }{\frac{J(\omega )}{\omega }}\mathrm{d}\omega.
\label{reorganizationenergy}    
\end{equation}
\textcolor{black}{This results in a semi-logarithmic sampling commonly used with an Ohmic behavior and an efficient description of the peaks of a structured spectral density. This sampling was also extended to use the correlation function obtained by semi-classical simulation directly \cite{Makri2017,Geva2020}. Methods based on an orthogonal polynomial strategy \cite{Chin2010,Prior2010,Prior2013,deVega-Wolf2015} are used more commonly in the linear chain mapping as discussed in Sec.\ref{sec:TWPmodes}. A unitary transformation towards the star model \cite{deVega-Wolf2015} would be possible but this approach has not been explored for the HEOM applications. }

\textcolor{black}{We now summarize the main lines of the adopted discretization procedure for D-HEOM. The approximate value of the renormalization energy with $M$ discrete modes is
\begin{equation}
{{\lambda }_{M}}=\frac{1}{2}\sum\limits_{m=1}^{M}{\frac{c_{m}^{D2}}{\omega _{m}^{D2}}}.    
\end{equation}
The contribution per mode is then
\begin{equation}
\varepsilon =\frac{1}{M}\int_{0}^{{{\omega }_{max }}}{\frac{J(\omega )}{\omega }}\mathrm{d}\omega =\frac{\pi {{\lambda }_{M}}}{M}    
\end{equation}
where $\omega_{max}$ is a cutoff frequency. The selected frequencies are determined by the constraint 
\begin{equation}
\frac{1}{\varepsilon }\int_{0}^{{{\omega }_{max}}}{\frac{J(\omega )}{\omega }}\mathrm{d}\omega =m-\frac{1}{2}   
\end{equation}
and the corresponding coefficient is given by
\begin{equation}
c_{m}^{D2}=\frac{2}{\pi }\frac{\omega _{m}^{{D}}J^D({{\omega }_{m}^D})}{{\rho }({{\omega^D }_{m}})}=\frac{2}{\pi }\omega _{m}^{D2}\varepsilon   
\end{equation}
where ${\rho }(\omega )=\sum\limits_{m}{\delta (\omega -{{\omega }_{m}^D})}$ is the state density.}

\textcolor{black}{Each discrete selected frequency $\omega_m^D$ selected on the real frequency axis leads to two artificial modes in Eq.(\ref{eq:Cdetexpan}) with complex frequencies $\gamma_k$ here becoming real positive and negative frequencies:
\begin{align}  \label{eq:gammaDM}
 {{\gamma }_{m1}^D}={{\omega }_{m}^D}  \nonumber  \\
 {{\gamma }_{m2}^D}=-{{\omega }_{m}^D}.
\end{align}
The corresponding coefficients are:
\begin{align}
\label{eq:alphaMD}
\alpha _{m1}^{{D}}=\frac{c_{m}^{D2}}{2{{\omega }_{m}^D}}\left( \frac{1}{{{e}^{\beta {{\omega }_{m}^D}}}-1} \right)  \nonumber  \\
{\alpha_{m2}^D}=\frac{c_{m}^{D2}}{2{{\omega }_{m}^D}} \left( \frac{{{e}^{\beta {{\omega }^D_{m}}}}}{{{e}^{\beta {{\omega }_{m}^D}}}-1} \right).
\end{align}
The relation between $\alpha _{m1}^{{D}}$ and $\alpha _{m2}^{{D}}$ satisfies the fluctuation-dissipation rule. This leads to write Eq.[\ref{eq:Cdetexpan}) as
\begin{equation}
 C(t)\approx \sum\limits_{j=1,2}{\sum\limits_{m=1}^{{{M}_{{}}}}{\alpha _{mj}^{D}}}{{e}^{i\gamma _{mj}^{D}t}}.   
\end{equation}
The total number of artificial modes in this discrete mapping used in D-HEOM is 
\begin{equation}
K^D_{HEOM}=2M 
\label{eq:Mmoddisc}
\end{equation}
with the same number of positive and negative frequencies in the star model as schematized in Fig.\ref{fig:fig3}(a).} 

 \section{Discrete modes in T-TEDOPA}
 \label{sec:TWPmodes}
 \subsection{Chain model from $J_{\beta}(\omega)$}
 \label{sec:TWPchainmodel}
The T-TEDOPA approach treats the dynamics of both the \textcolor{black}{open quantum system} and the environmental selected modes in the Hilbert space at a finite temperature by the outstanding T-TEDOPA method \cite{Tamascelli2019,Tamascelli2020,Dunnett2021,Riva2023}. This very efficient strategy replaces the initial thermalized mixed state by the pure state of the environment at 0 K. The initial total density operator [Eq.(\ref{rhoini})] assumed to be factorized into the system density operator and the thermal equilibrium bath density operator is superseded in the T-TEDOPA algorithm by 
\begin{equation}
 {{\rho }_{tot}}(t=0)={{\rho }_{S}}(t=0)\otimes \left| 0...0 \right\rangle \left\langle  0...0 \right|   
\end{equation}
where $\left| 0...0 \right\rangle$ is the pure state of the environment with all the modes in their ground state. The equivalence is ensured by considering an extended frequency range, comprising negative frequencies and by sampling not the usual spectral density $J(\omega)$ on the positive frequency axis as done in Sec.\ref{sec:Dmodes} but the temperature dependent $J_{\beta}(\omega)$ [Eq.(\ref{eq:Jbeta})] on the extended axis. For finite temperatures, $J_{\beta}(\omega)$ is asymmetrical with
\begin{equation}
 J_{\beta}(-\omega)=\exp(-\beta \omega) J_{\beta}(\omega).   
\end{equation}
This allows to fix a minimum cut-off frequency $\omega_{min}$ on the frequency negative axis and in principle a lower number of negative frequencies. 

\begin{figure}
    \centering
    \includegraphics [width =1.\columnwidth, height =6 cm]{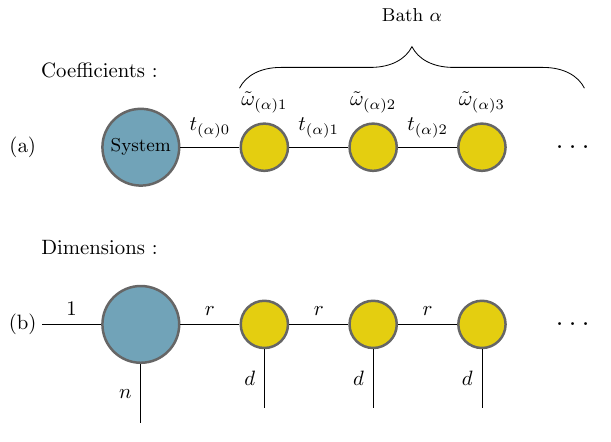}
    \caption{ (a) T-TEDOPA tensor configuration for the example of a unique bath $\alpha$ interacting with a system. The coefficients are obtained with the orthogonal polynomial method.  (b) Dimensions of the MPS. Chain tensors of local Fock space $d$ are bounded with the bond dimension $r$. Only the first chain tensor is interacting with the system of dimension $n$.}
    \label{fig:fig4}
\end{figure}

The second main difference consists in working with a linear chain model \cite{Chin2010,Prior2013,Plenio2019,Dunnett2021} as schematized in Fig.\ref{fig:fig3}(b). This procedure is based on the theory of orthogonal polynomials. The bath Hamiltonian of a star model [Eq.(\ref{eq:HB})] becomes after unitary transformation of the modes the transformed Hamiltonian of the chain model:
\textcolor{black}{
\begin{align}
{\tilde{H}_{B}}= \frac{1}{2} \sum\nolimits_{\alpha }^{{{N}_{bath}}}{\sum\nolimits_{j}^{{{N}_{\alpha }}}{\tilde{\omega} _{\alpha j}^{2}}x_{\alpha j}^{2}} \nonumber \\
+\sum\nolimits_{\alpha }^{{{N}_{bath}}}\sum\nolimits_{j}^{{{N}_{\alpha }}}t_{\alpha j}x_{\alpha j}x_{\alpha (j+1)}
\label{eq:HBchain}
\end{align}
with $\tilde{\omega}_{\alpha j}$ the frequency of the $j^\text{th}$ chain mode $x_{\alpha j}$ and $t_{\alpha j}$ its nearest-neighbour coupling coefficient.  The coupling between the system operators $S_{\alpha}$ and the bath involves the first mode $x_{\alpha 1}$ of the chain only
\begin{equation}
\tilde{H}_{SB} = \sum\nolimits_{\alpha }^{{{N}_{bath}}} S_{\alpha} \otimes t_{\alpha 0} x_{\alpha 1}.    
\end{equation}
} We refer the reader to other works detailing the theory and how to obtain the frequencies $\tilde{\omega}_{\alpha j}$ and the coefficients $t_{\alpha j}$ from $J_\beta(\omega)$ \cite{Chin2010,Prior2013}. \textcolor{black}{The general configuration of the chain model is shown in Fig.\ref{fig:fig4}(a).}
  \textcolor{black}{Ensuring convergence of $C(t)$ within a finite timescale as suggested recently to systematically coarse grain an environment \cite{Plenio2024} is a relevant condition for truncating the chain. However, due to the linear form of the environment, excitation emitted by the system will diffuse into the chain. The length of the chain must be set up in order to avoid the return of the excitation to the system, which would perturb it without physical meaning. The total number of artificial modes in T-TEDOPA is illustrated in Fig.\ref{fig:fig4}(b) and is given by the optimized truncated chain of $N_{ch}$ modes:
\begin{equation}
K^D_{TEDOPA}=N_{ch}
\label{eq:MmodeTEDOPA}    
\end{equation}
}

The T-TEDOPA method propagates the full many body system-environment wave function in the MPS format that is particularly efficient for discretized and linearized baths. The system and each mode of the chain is a site or core of the MPS. The bond dimension usually denoted $D$ in this MPS context \cite{Schröder2019} will be called $r$ here as in the HEOM applications \cite{Shi2020}. By assuming a constant bond dimension $r$, $N_{ch}$ modes and a dimension $d$ for each Fock space, the storage requires 
\begin{equation}
N_{TEDOPA}=N_{ch}dr^2 
\label{eq:NTEDOPA}
\end{equation}
complex elements. Note that $d$ corresponds to the level $L$ in HEOM ($d=L+1$). The chain tensors of local Fock spaces is schematized in Fig.\ref{fig:fig4}(b).

 \subsection{Discrete $\textit{ab}$ $\textit{initio}$ vibrators}
 \label{sec:TWPabinitio}
 The discrete modes of the initial star model before the transformation into the chain model may be the molecular vibrators of a LVC model calibrated from $\textit{ab}$ $\textit{initio}$ data. The discrete coefficients  $c_m$  [Eq.(\ref{eq:Jdiscret})] are then obtained from the gradients of the energies or of the interstate coupling at the reference geometry of the baths at equilibrium. Dynamics at a finite temperature then involve modified coefficients $c_m^{(\beta)}$ by using:
 \begin{equation}
 c_m^{(\beta)2}=c_m^2 \left( 1+n_{\beta}(\omega_m) \right)
  \label{eq:gmbeta}   
 \end{equation}
 The ensemble of modes is extended to include some negative frequencies ($\omega_m$ and $- \omega_m$) according to the value of the Bose function on the negative axis. The distribution of the frequencies is therefore asymmetrical. The oscillator chain is then built by the procedure based on the orthogonal polynomials \cite{Chin2010,Schröder2019}.
 
\section{Numerical methods}
\label{sec:nummethod}
The numerical investigation is carried out with HEOM in conventional or MPS implementation and with T-TEDOPA. HEOM  [Eqs.(\ref{eq:eqcouHEOM})] are solved with each kind of artificial decay modes presented in Sec.\ref{sec:modes}. The detailed equations for each case are summarized in the Supplementary Material. The equations with standard storage of the matrices are solved using the Runge–Kutta 4 (RK4), Cash–Karp (RK4-5) with adaptative step-size or Arnoldi algorithms. Some dynamics may be driven only with the MPS format \cite{Shi2018,Shi2020,Shi2021,Borrelli2021,Ke2023,Mangaud2023} for which we use the projector-splitting KSL scheme \citep{Lubich2014,Lubich2015,Oseledets2016,Batista2022} implemented in the ttpy package (\verb|tt.ksl.ksl|) \cite{ttpy}. The method is based on the dynamical low-rank approximation which is equivalent to the Dirac-Frenkel time-dependent variational principle. It consists in using an approximate low-rank tensor with fixed ranks instead of getting a solution with a high rank tensor and then truncate it with singular value decomposition (SVD). To reach this goal, the derivative of the approximate low-rank tensor is obtained by projecting the derivative of the tensor on the tangent space of the approximate low-rank tensor at its current position. Time-integration is then obtained by a splitting scheme (second order in this work) of the projector  \citep{Lubich2014,Lubich2015,Oseledets2016,Batista2022}. An adaptive bond dimension $r$ is necessary during the propagation as proposed in Refs.\cite{DunnettChin2021,Dolgov19}. We have adopted a mixed strategy. The standard Runge-Kutta integrator (written with TT algebra available with the \verb|ttpyl| package) is run after 10 time steps to allow the increase of the rank during the propagation. 

In the T-TEDOPA method, the MPS is  propagated using one-site time-dependent variational method (1TDVP). With this time-evolution method,  the most costly operation has a complexity of $\mathcal{O} \left( r^2 d^2 w^2 + r^3 d w + r^3 d^2 \right)$ with $w$ the bond dimension of the time-evolution operator \cite{li_numerical_2020, Dunnett2021}. Although not used in these simulations, 1TDVP includes an important suitability for bosonic environments as it also allows an adaptive bond dimension \cite{DunnettChin2021,Dolgov19}. All of the T-TEDOPA simulations were carried out using the Julia package MPSDynamics \cite{MPSDynamics}.

\section{Illustrative simulations}
\subsection{Pure dephasing model}
\label{sec:PD}
\begin{figure}
    \centering
    \includegraphics [width =0.8\columnwidth, height = 5.5 cm]{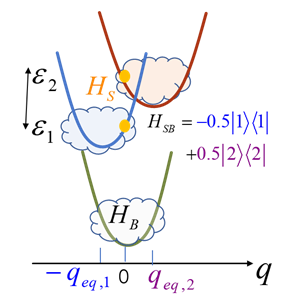}
    \caption{Spin-Boson model for a pure dephasing case. The system Hamiltonian  $H_S$ is expressed in the eigen basis set of the local electronic Hamiltonian. The two energies correspond to a Franck-Condon transition towards uncoupled excited electronic states. The equilibrium position of each oscillator in $H_B$ is set $q = 0$ where $q$ is a generic vibrational coordinate. The $H_{SB}$ coupling is due to the displacement of the oscillator equilibrium position in each excited state inducing an energy gradient. In this example, the baths are anti-correlated since the displacements are in opposite directions. }
    \label{fig:fig5}
\end{figure}

We consider a pure dephasing spin-Boson model (SB) for which an analytical solution exists \cite{Breuer2002}, allowing for straightforward calibration of the number of artificial modes. The model is schematized in Fig.\ref{fig:fig5}. The system Hamiltonian is $H_S=(\Delta E /2) \sigma_z$ where $\sigma_z$ is a Pauli matrix. $\Delta E= (\epsilon_2 - \epsilon_1)= 0.002$ Hartree. The system is diagonally coupled to the bath with a coupling operator $S= \sigma_z/2$. This model  corresponds to two excited electronic states in which the equilibrium positions of the oscillators are displaced in opposite directions with respect to that of the ground state bath, indicating anti-correlation between the baths \cite{Schulten2011}. 

The decoherence function $D(t)=|\rho_{S_{12}}(t)|$ is the modulus of the off-diagonal element of the reduced system density matrix. It evolves as:
\begin{equation}
\frac{D(t)}{D(0)}= {{e}^{-\int_{0}^{\infty }{\frac{J(\omega )}{{{\omega }^{2}}}\coth \left( \frac{\beta \omega }{2} \right)(1-\cos (\omega t))\mathrm{d}\omega }}}.
\label{eq:coheana}
\end{equation}
After the mapping with $M$ discrete modes, the decoherence function becomes:
\begin{equation}
\frac{D(t)}{D(0)}={{e}^{-\sum\limits_{j=1}^{M}{\frac{c_{j}^{2}}{\omega _{j}^{3}}}\coth \left( \frac{\beta {{\omega }_{j}}}{2} \right)(1-\cos ({{\omega }_{j}}t))}}.
\label{eq:cohedis}
\end{equation}

\begin{figure}
    \centering
    \includegraphics [width =1.\columnwidth]{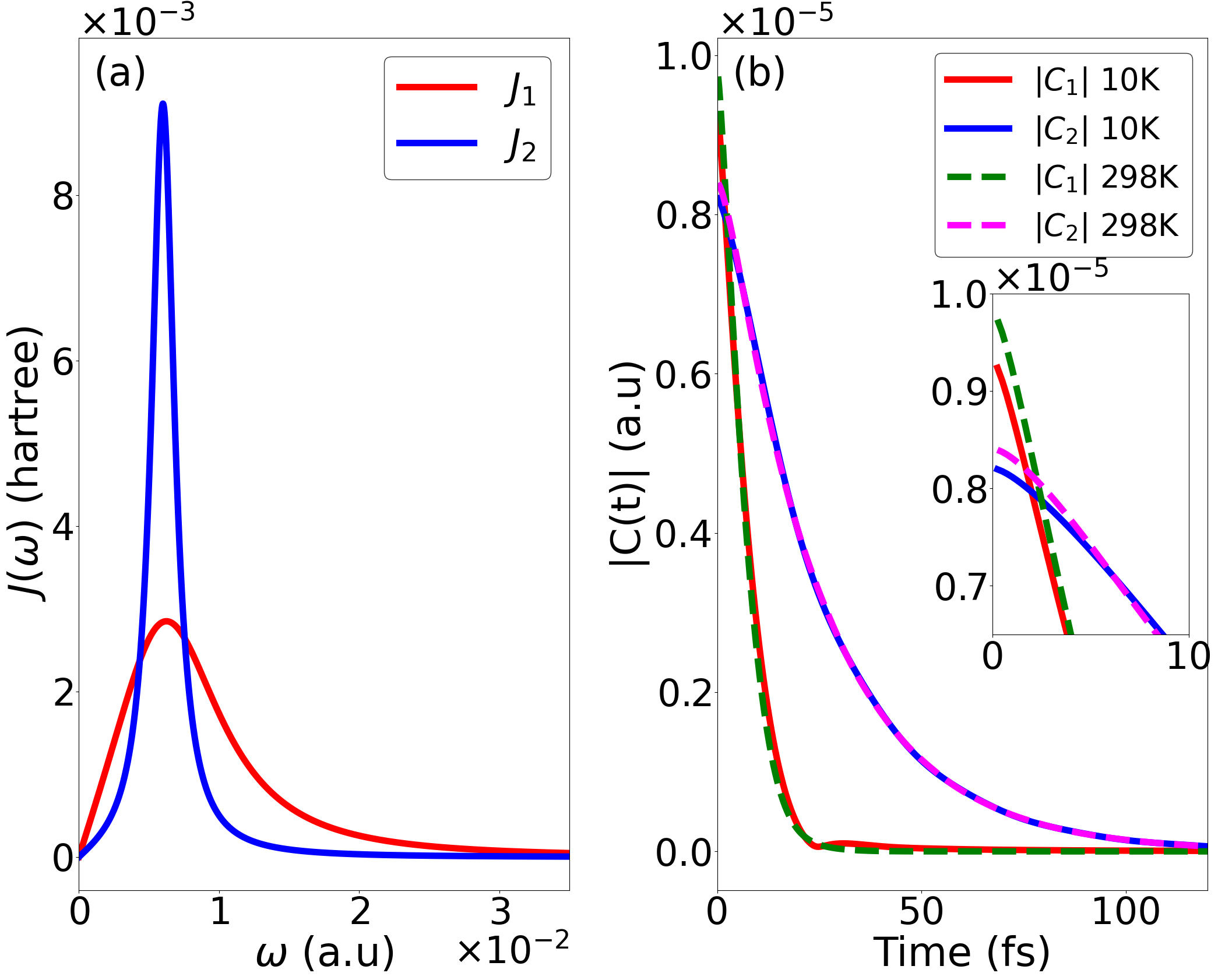}
    \caption{(a) Spectral densities $J_1(\omega)$ and $J_2(\omega)$ used in the pure dephasing SB model. The parameters ${p_l, \Omega_l, \Gamma_l}$ of the Tannor-Meier Lorentzian functions [Eq.(\ref{eq:JTM})] are given in the Supplementary Material. (b) Modulus of the corresponding correlation functions [Eq.(\ref{eq:Cdet})] at 10 K and 298 K.}
    \label{fig:fig6}
\end{figure}

The spectral densities of the two examples are displayed in Fig.\ref{fig:fig6}(a). They are Tannor-Meier functions [Eq.(\ref{eq:JTM})] with similar renormalization energy ($\lambda_1=1.64 \times 10^{-3}$ Hartree for $J_1(\omega)$ and $\lambda_2=1.49 \times 10^{-3}$ Hartree $J_2(\omega)$) giving a ratio $\lambda/\Delta E$ around 0.75 that corresponds to a strong coupling regime. The parameters $p_l$, $\Omega_l$ and $\Gamma_l$ [Eq.(\ref{eq:JTM})] are gathered in the Supplementary Material. The corresponding correlation functions are given in Fig.\ref{fig:fig6}(b) for $T =10$ K and $T = 298$ K. 

\begin{figure*}
    \centering
    \includegraphics [width =2.\columnwidth]{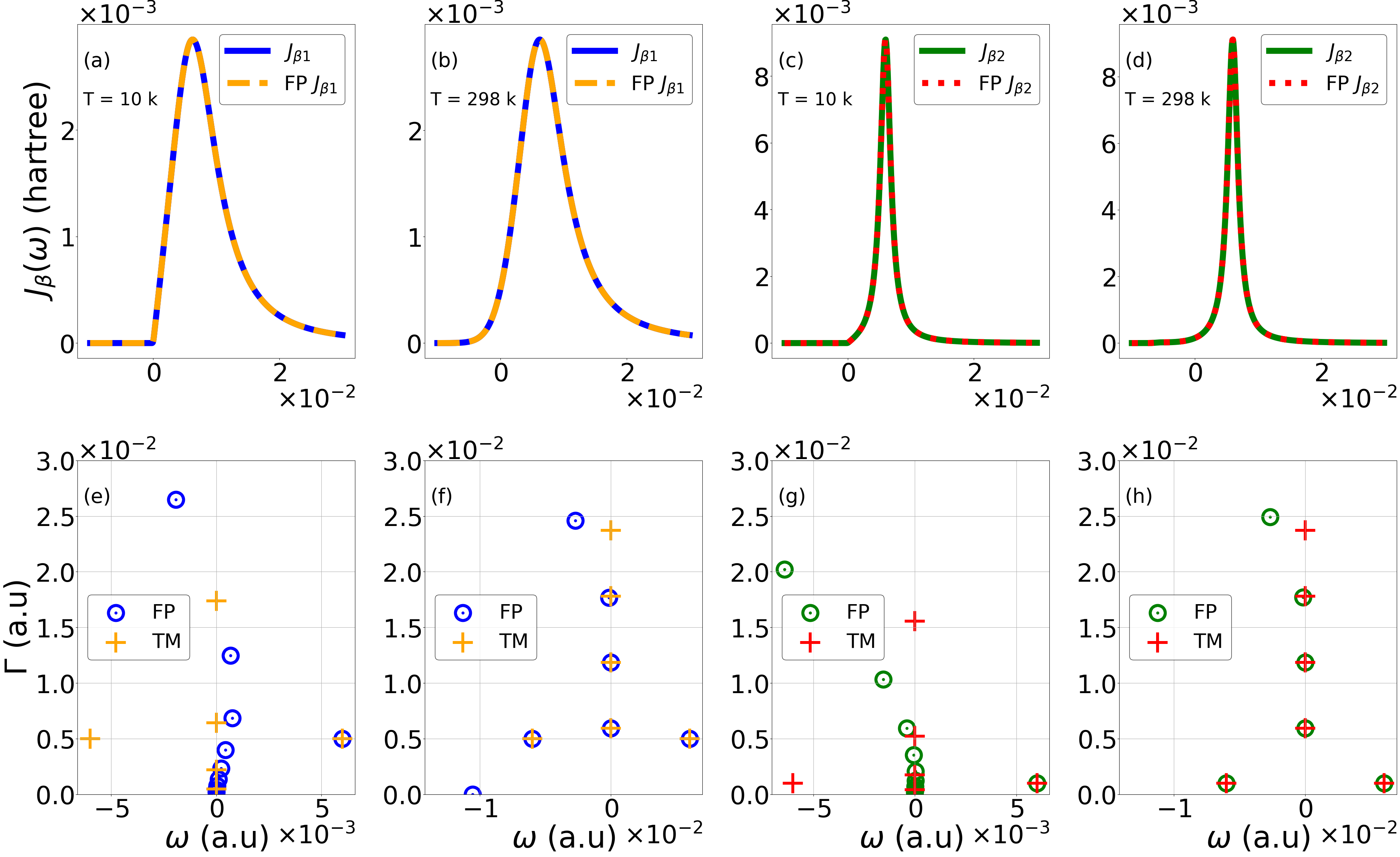}
    \caption{ Temperature dependent spectral densities $J_{\beta 1}(\omega)$ (a), (b) and $J_{\beta 2}(\omega)$ (c), (d) at 10 K and 298 K in solid lines [Eq.(\ref{eq:Jbeta})] and their fits by the barycentric expansion [Eq.(\ref{eq.AAA})] in dashed lines. The lower panels give the corresponding free poles (FP) (full circles) and the poles of the Tannor-Meier (TM) parametrization [Eq.(\ref{eq:JTM})] in the upper half-plane $\Gamma > 0$ (crosses). The four decay rates $ \gamma_k$ used in the fitting of the Matsubara contribution to $C(t)$ [Eqs.(\ref{eq:CMdet}), (\ref{eq:CMdetfit})] are also indicated (crosses on the imaginary axis). \textcolor{black}{For comparison, the spread in frequencies is $[-0.001, 0.018]$ a.u. at 10 K and $[-0.003, 0.018]$ a.u. at 298 K for T-TEDOPA. It is $[-0.014, 0.014]$ a.u. in D-HEOM.} }
    \label{fig:fig7}
\end{figure*}

Figure \ref{fig:fig7} shows the fitting of $J_{\beta 1}(\omega)$ and $J_{\beta 2}(\omega)$ by the barycentric expansion for $T =10$ K and $T = 298$ K. The free poles (FP) in the upper half-plane are also given. We add the poles of the Tannor-Meier function of $J_1(\omega)$ or $J_2(\omega)$ [Eq.(\ref{eq:JTM})] and the four $ \gamma_{k}$ values used for fitting the Matsubara contribution to the correlation function [Eqs.(\ref{eq:CMdet}), (\ref{eq:CMdetfit})]. They are represented by cross symbols.  We use four real exponential functions (the parameters are given in the Supplemental Material). 

\begin{figure}
    \centering
    \includegraphics [width =1.\columnwidth]{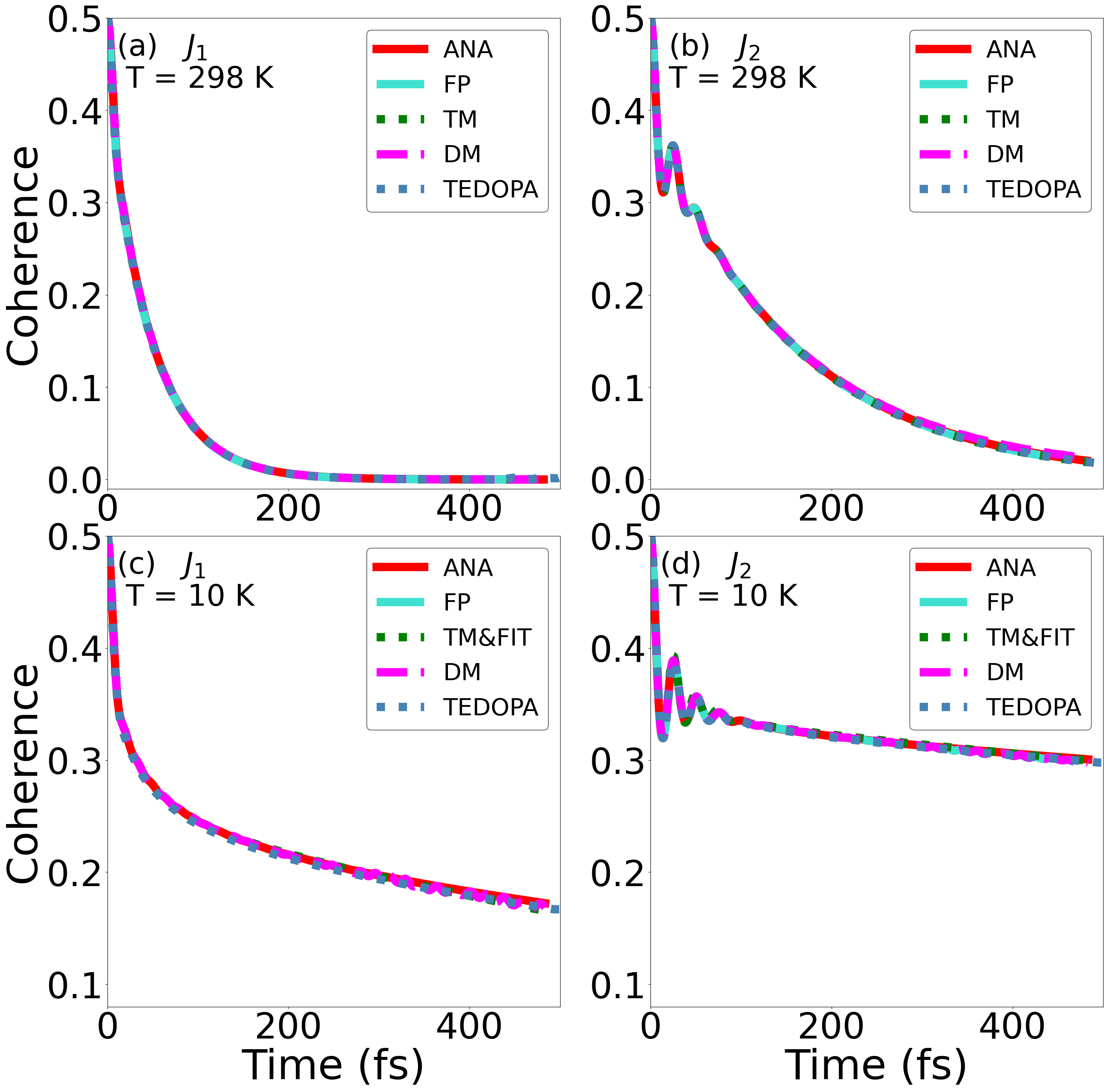}
    \caption{Decoherence function $D(t)=|\rho_{S_{12}}(t)|$ of the pure dephasing SB model with the spectral densities  $J_{1}(\omega$) (panels (a) and (c)) and $J_{2}(\omega$) (panels (b) and (d)). The benchmark is the analytical expression [Eq.(\ref{eq:coheana})] (ANA). The number of discrete modes (DM) is fixed by converging the discrete mapping [Eq.(\ref{eq:cohedis})]. A very good accuracy is reached for the HEOM methods (FP-HEOM, TM$\&$FIT, D-HEOM) and for the T-TEDOPA simulation at high temperature (a) and (b) and at low temperature (c) and (d).} 
    \label{fig:fig8}
\end{figure}

 Figure \ref{fig:fig8} gives the decoherence function $D(t)$ for $J_1(\omega)$ and $J_2(\omega)$ at $T$ = 298 K in panels (a) and (b) respectively and at $T$ = 10 K in panels (c) and (d) computed by the three HEOM methods (FP-HEOM, TM\&FIT, undamped discrete modes D-HEOM) and by T-TEDOPA. The analytical expression [Eq.(\ref{eq:coheana})] (ANA) is the reference to check the convergence. $J_1$ leads to an overdamped situation with a monotonous decay. The decoherence in the $J_2$ example exhibits small oscillations, typical of a transitory non-Markovian behavior. Indeed, the decoherence function of the SB model is closely related with a measure of non-Markovianity given by the volume of accessible states for the system. Any bump in the volume or in this case in the decoherence function is a signature on non-Markovianity, indicating a transitory flow back from the reservoir to the system \cite{Paternostro2013,Mangaud_2018,Amati2024}. This is closely linked to the oscillation of the collective bath mode \cite{ChinChevet2019}.

\begin{figure}
    \centering
    \includegraphics [width =1.\columnwidth,height = 6 cm]{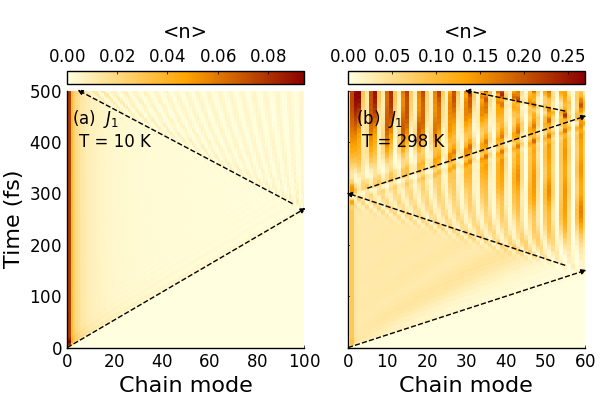}
    \caption{Dynamics of the averaged chain mode populations $\langle n \rangle$. The system is coupled to the first chain mode. Dashed black arrows are guide eyes to follow the first excitation propagation. (a) $J_1$ chain at $T = 10$ K in which the length is regulated with the first excitation reaching the system. (b) $J_1$ chain at $T =298$ K where the first back and forth propagation does not perturb the system significantly, which allows a shorter chain.} 
    \label{fig:fig9}
\end{figure}

 \textcolor{black}{In both HEOM and T-TEDOPA strategies, we determine the minimum number of modes to reach convergence of the coherence for a timescale $\tau = 500$ fs. The number of modes can be difficult to compare. The criterion of the convergence of the bath correlation function for the selected timescale $\tau$ must be satisfied in each method. However, the increase rate of the number of modes $K$ with $\tau$ is not necessarily the same. It depends on the characteristics of the propagation along the chain in T-TEDOPA. For very short $\tau$, $K$ may be smaller in T-TEDOPA than in HEOM but this may change for longer simulation times. The comparison is here for a timescale $\tau = 500$ fs (see Fig.\ref{fig:fig8}) and it does not permit a generalization to many others situations. This interesting question could be examined in a further work.} An illustration of the convergence of the coherence for D-HEOM and T-TEDOPA is included in the Supplementary Material. Table \ref{tab:pure_dephasing_modes} gives the number of modes and the number of complex elements required in each numerical simulation. In the D-HEOM case, the number of modes is larger for the more non-Markovian example with $J_2(\omega)$ and it increases with decreasing temperature. In the T-TEDOPA simulation, the temperature dependent spectral densities $J_{\beta}(\omega)$ are asymmetric with a negative cutoff frequency of -0.001 Hartree at 10 K and -0.003 Hartree at 298 K. The positive cutoff is 0.018 Hartree for every case. Concerning the $J_2(\omega)$ spectral density, the number of modes $N_{ch}$ is expected to increase with $T$. However, the minor difference in the negative cutoff frequencies for both temperatures leads to very similar chains. The propagation of the excitation along the chain depending on the length, the difference at 10 K and 298 K is negligible for $J_2(\omega)$ \cite{Riva2023, DunnettTemp2021}. \\ 
\indent Although the situation is similar concerning negative cutoff frequencies, the Ohmic shape at low frequency of $J_1(\omega)$ results in a different behavior. Figure \ref{fig:fig9} shows the dynamics of the average chain mode population $\langle n \rangle$ for the whole chain at both temperatures. One can see that excitation is injected in the chain at each time step. At $T =10$ K, the chain length is based on the return of the first excitation to the system. Due to the specific Ohmic shape of $J_1(\omega)$, excitation can reach the system (coupled to the chain mode 1) at $T =298$ K without much effects up to one point. Such process would instead lead to discrepancies against the analytical result within our convergence criteria for the  $T =10$ K case. Fringes are also visible in Fig.\ref{fig:fig9}, due to excitation crossings in the chain. The result concerning $T =298$ K means that excitation injected in the chain at early time is no longer important for the system once it reaches it. This illustrates an interesting way to circumvent the common sampling criterion for such environments, usually based on avoiding excitation returns as in the $T = 10$ K example. It might also play a role for Markovian closures \cite{NelerMarkovClosure2022}. \textcolor{black}{The observation made in this example may be model dependent. Our example only represents an interesting observation for this specific case which does not imply generality. We would have to look at several types of spectral density and system-bath couplings to result in something global.
}

\begin{table*}[ht]
    \centering
    \textcolor{black}{
    \begin{tabular}{lcccccc}
        \toprule
        \toprule
        & \multicolumn{5}{c}{Method} \\
        \cmidrule{3-7}
        T & $J(\omega)$ & ~ & D-HEOM & TM\&FIT-HEOM & FP-HEOM & T-TEDOPA \\
        \midrule
        10 K & $J_1$  & ~ & ~ & \\
         ~  & ~ & $K$ & 80 & 2 + 4 & 20  & 100 \\
         ~  & ~ & $N^{st}_{HEOM}$ & 367,524 & 336 & 7,084  & ~ \\
        ~  & ~ & $N^{MPS}_{HEOM}$ & 284,640 & 18,240 & 68,840  & ~ \\
         ~  & ~ & $N_{TEDOPA}$ & ~ & ~ & ~ & 3,600 \\
         ~ & $J_2$  & ~ & ~ & \\
        ~  & ~ & $K$  & 260 & 2 + 4 & 20 & 110 \\
         ~  & ~ & $N^{st}_{HEOM}$ & 13,415,584 & 336 & 7,084  & ~ \\
        ~  & ~ & $N^{MPS}_{HEOM}$ & 932,640 & 18,840 & 68,840  & ~ \\
         ~  & ~ & $N_{TEDOPA}$ & ~ & ~ & ~ & 3,960 \\
         \midrule
        298 K & $J_1$  & ~ & ~ & \\ 
        ~  & ~ &$K$ & 80 & 2 + 4 & 20 & 60 \\
        ~  & ~ & $N^{st}_{HEOM}$ & 367,524 & 336 & 7,084  & ~ \\
        ~  & ~ & $N^{MPS}_{HEOM}$ & 284,640 & 18,840 & 68,840  & ~ \\
         ~  & ~ & $N_{TEDOPA}$ & ~ & ~ & ~ & 2,160 \\
         ~ & $J_2$  & ~ & ~ & \\
        ~  & ~ & $K$ & 200 & 2 + 4 & 20 & 110 \\
         ~  & ~ & $N^{st}_{HEOM}$ & 6,351,944 & 336 & 7,084  & ~ \\
        ~  & ~ & $N^{MPS}_{HEOM}$ & 716,640 & 18,840 & 68,840 & ~ \\
         ~  & ~ & $N_{TEDOPA}$ & ~ & ~ & ~ & 3,960 \\
        \bottomrule
        \bottomrule 
    \end{tabular}}
    \caption{\textcolor{black}{ \label{tab:pure_dephasing_modes} Number of artificial modes $K$, (Eqs. (\ref{eq:Mmoddisc}) for D-HEOM, (\ref{eq:MTMFIT}) for TM$\&$FIT-HEOM,  (\ref{eq:MmodesFP}) for FP-HEOM and (\ref{eq:MmodeTEDOPA}) for T-TEDOPA) used in the simulation of the pure dephasing qubit. The number of estimated complex elements in the storage is indicated (Eqs.(\ref{eq:NHEOMst}) for the HEOM standard implementation without ADOs filtering, (\ref{eq:NHEOMMPS}) for the MPS representation, and (\ref{eq:NTEDOPA}) for T-TEDOPA). The HEOM level is $L$ = 3.  In the simulations made in the MPS format, the maximum bond dimension is $r_{max} = 30$ (it should be noted that $r$ may be smaller than $r_{max}$ during the propagation) and the tolerance is  10$^{-10}$. In the T-TEDOPA simulation, the Fock space has dimension $d = 4$ and bond dimension $r = 3$.} }
\end{table*}
\textcolor{black}{
All the simulations merge with the analytical solution each having its distinct advantages and disadvantages, particularly at low temperatures. (i) At low temperature, the number of Matsubara terms [Eq.(\ref{eq:CMdet})] is too high so that the TM parametrization is possible only by fitting of the Matsubara contribution [Eq.(\ref{eq:CMdetfit})] \cite{Lambert2019}. This procedure is particularly effective. The FP-HEOM strategy also reveals its efficiency in this application since few poles of $J_{\beta}(\omega)$, only 2 $\times$ 10 terms, are necessary for low temperature
. The fact that TM$\&$FIT is particularly interesting here is due to the shape of the spectral density involving a single peak. It is obvious that highly structured densities with greater renormalization energy requires more TM functions and high hierarchy level $L$ leading to a larger number of ADOs. (ii) The discrete mapping used in D-HEOM involves a larger number of modes making implementation with standard encoding prohibitive. Only the MPS format is efficient in this example. T-TEDOPA working with wave functions instead of density matrices requires fewer resources and confirms its outstanding efficiency when compared to D-HEOM. }

\subsection{Two-bath model}
\label{sec:twobath}
\begin{figure}
    \centering
    \includegraphics [width =0.9\columnwidth, height = 6 cm]{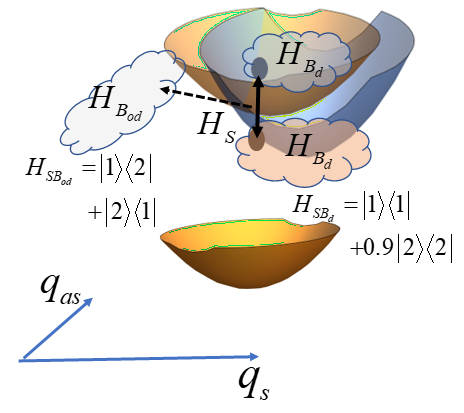}
    \caption{Spin-Boson model of a conical intersection between two excited states of a molecular system. Two correlated baths ($H_{B_d}$) are diagonally coupled to the local electronic system Hamiltonian and one bath ($H_{B_{od}}$) is off-diagonally coupled to the inter-state coupling. The reference equilibrium position of the bath oscillators is that of the ground state. ${{H}_{S}}={{\varepsilon }_{1}}\left| 1 \right\rangle \left\langle  1 \right|+{{\varepsilon }_{2}}\left| 2 \right\rangle \left\langle  2 \right|$ corresponds to the diabatic energies at the Franck-Condon transition where the  inter-state coupling vanishes. The tuning $H_{B_d}$ and coupling $H_{B_{od}}$ baths gather the symmetrical or antisymmetric vibrational modes denoted $q_s$ and $q_{as}$ respectively.}
    \label{fig:fig10}
\end{figure}

We now consider a more demanding simulation with a SB model in which the system is coupled diagonally and off-diagonally to baths in a strong coupling regime.  In chemical physics, this model describes a conical intersection between two excited states \cite{Domcke2004,Domcke2011} by partitioning a local electronic Hamiltonian coupled to two manifolds of vibrational modes of different symmetries (symmetrical or antisymmetric with respect to the symmetry group at the reference equilibrium position of the ground state). A generic scheme is given in Fig.\ref{fig:fig10}. ${{H}_{S}}={{\varepsilon }_{1}}\left| 1 \right\rangle \left\langle  1 \right|+{{\varepsilon }_{2}}\left| 2 \right\rangle \left\langle  2 \right|$ corresponds to the electronic diabatic energies and inter-state coupling at the Franck-Condon transition. As recalled below, a diabatic representation is defined up to a unitary transformation. We work in the basis corresponding to the eigenstates at the Franck-Condon geometry where the electronic coupling vanishes. Such a partition has already been proposed \cite{Thorwart2016,MDL_19} and leads to a highly non-Markovian behavior requiring a high HEOM hierarchy level. Baths diagonally coupled to the system make fluctuate the energies and are sometimes called tuning baths. They gather the symmetrical modes. To reduce the number of baths, we assume that these diagonal baths are correlated \cite{Schulten2011}, i.e. the equilibrium positions of the excited states are displaced in the same direction from the reference equilibrium position. One may then consider a single spectral density $J_d(\omega)$ (with subscript $d$ for diagonal) and the corresponding system-bath coupling operator then writes $S_{{B}_{d}}=\left| 1 \right\rangle \left\langle  1 \right|+\alpha \left| 2 \right\rangle \left\langle  2 \right|$. The other bath (also called the coupling bath) is formed by the antisymmetric modes. It is coupled to the off-diagonal elements by the operator ${{S_{{B}_{od}}}}=\left| 1 \right\rangle \left\langle  2 \right|+\left| 2 \right\rangle \left\langle  1 \right|$.  The interstate coupling cancels at the reference point and takes a finite value when the coupling bath modes oscillate. The corresponding spectral density is $J_{od}(\omega)$ (with subscript $od$ for off-diagonal). The model is calibrated from the  \textit{ab} \textit{initio} data LVC model of a phenylene ethynylene dimer (1,3-bis(phenylethynyl)benzene) computed by Lasorne \textit{et} \textit{al}. \cite{Ho2019,Jaouadi2022,Lasorne2023}. The simulations with the intermolecular vibrational modes (\ref{sec:TWPabinitio}) use the parameters of the LVC model (gradients of the energies and of the interstate coupling at the minimum of the ground electronic state) given in the Supplemental Information of Ref.\cite{Jaouadi2022}. The tuning and coupling baths gather 35 and 34 symmetrical ($A_1$) or antisymmetric ($B_2$) respectively. In order to account for a surrounding, the delta distribution in the spectral density [Eq.(\ref{eq:Jdiscret})] is broadened by a Lorentzian smoothing function
\begin{equation}
\delta (\omega -{{\omega }_{m}})\to \frac{1}{\pi }\frac{\Gamma  }{{{(\omega -{{\omega }_{m}})}^{2}}+\Gamma^2}    
\end{equation}
leading to a continuous spectral density with the same renormalization energy. In this work, we used $\Gamma = 160$ cm$^{-1}$. The excited manifold is not coupled to the ground state by inter-state coupling but only radiatively. 

\begin{figure}
    \centering
    \includegraphics [width =1.\columnwidth]{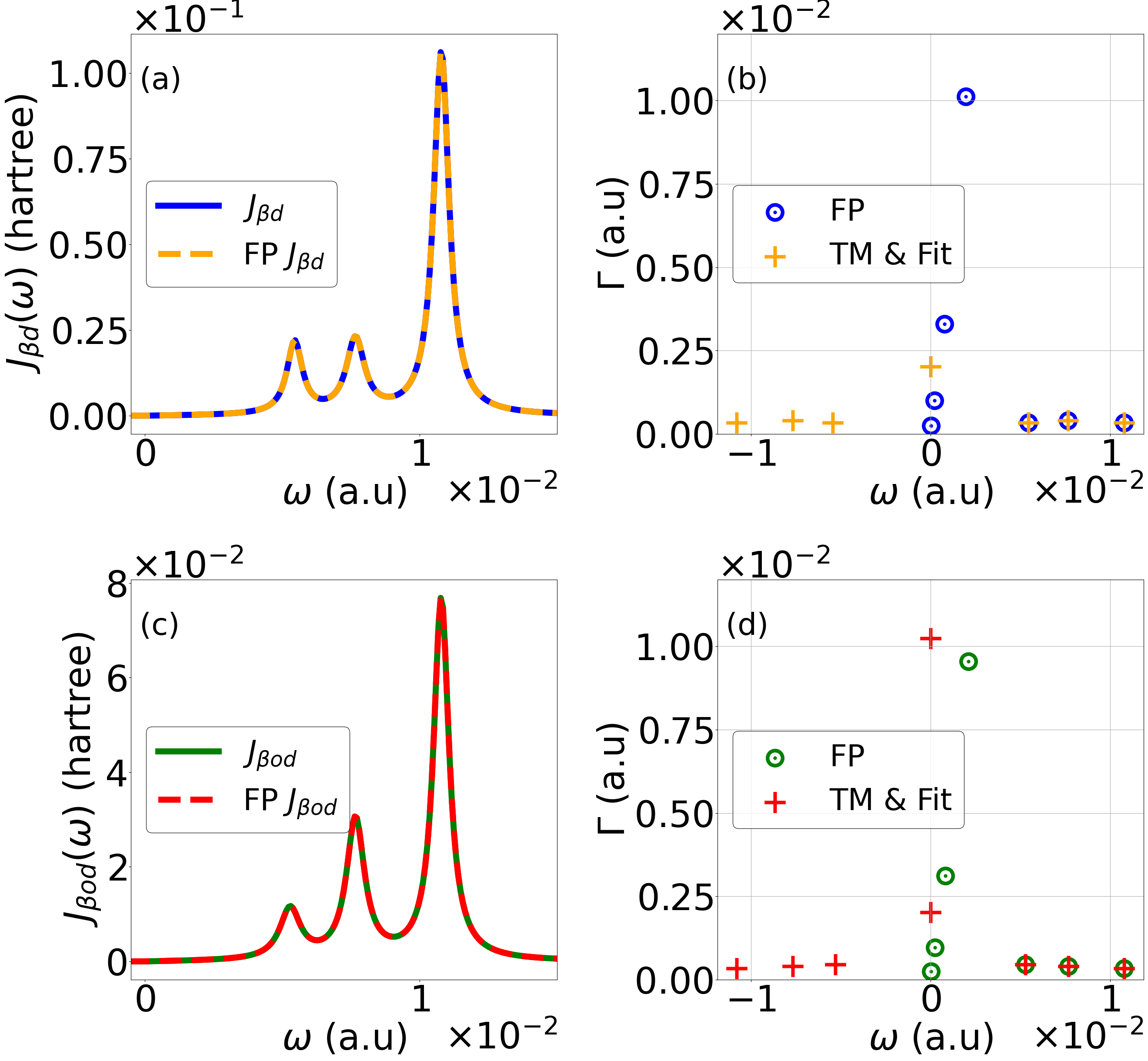}
    \caption{Temperature dependent spectral densities $J_{\beta d}(\omega)$ (a) and $J_{\beta od}(\omega)$ (c) and there fit by the barycentric expansion [Eq.(\ref{eq.AAA})] at 10 K. Panels (b) and (d) give the corresponding free poles (FP) (circles), the poles of the TM parametrization [Eq.(\ref{eq:JTM})] in the upper half-plane $\Gamma > 0$ (crosses) and the two decay rates $\gamma_k$ used in the fitting of the Matsubara contribution to $C(t)$ [Eq.(\ref{eq:CMdetfit})] (crosses on the imaginary axis). }
    \label{fig:fig11}
\end{figure}

Figure \ref{fig:fig11} shows the $J_{\beta d}(\omega)$ (a) and $J_{\beta od}(\omega)$ (c) spectral densities of the tuning $H_{Bd}$ and coupling $H_{Bod}$ baths at 10 K. The poles (circles) of the barycentric expansions [Eq.(\ref{eq.AAA})] in the upper half-plane $\Gamma > 0$ are displayed in panels (b) and (d) respectively. To compare with the TM$\&$FIT method, $J_d(\omega)$ and $J_{od}(\omega)$ are fitted by three TM Lorentzian functions. The poles of these TM functions in the upper half-plane and the rates used in the Matsubara fitting procedure [Eqs.(\ref{eq:CMdet}) and (\ref{eq:CMdetfit})] are superimposed (crosses) in Fig.\ref{fig:fig11}. The parameters [Eqs.(\ref{eq:JTM}) and (\ref{eq:CMdetfit})] are gathered in the Supplemental Material.

\begin{figure}
    \centering
    \includegraphics [width =1.\columnwidth]{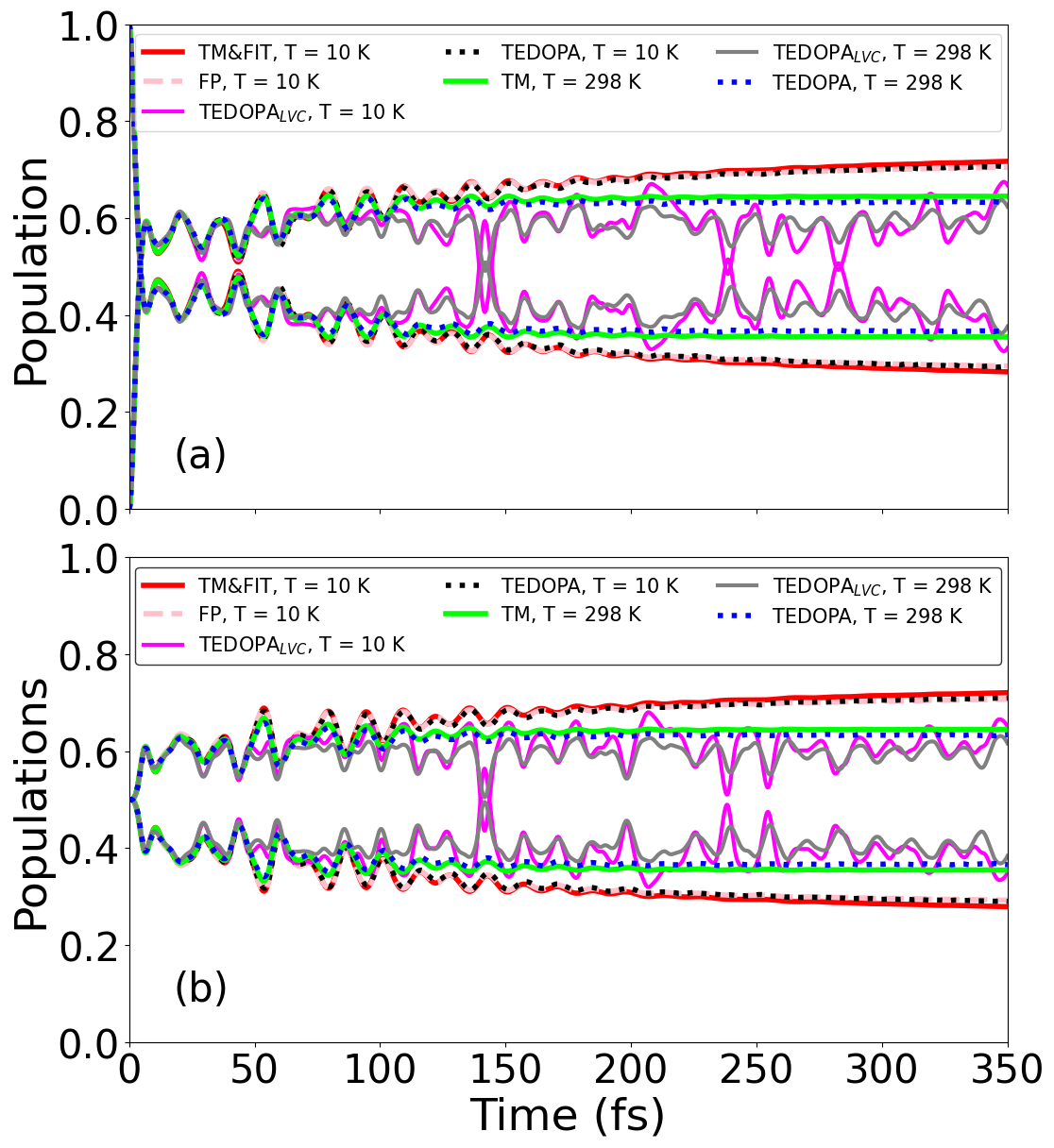}
    \caption{ Population evolution at 10 K and 298 K in the two-bath model computed by HEOM (FP, TM\&FIT), T-TEDOPA with discrete modes extracted from continuous spectral densities or using the $\textit{ab}$ $\textit{initio}$ LVC parameters (T-TEDOPA$_{LVC}$). (a) The initial condition is the upper excited state $\left| 2 \right\rangle$, (b) preparation in a superposition of the two excited states $(\left| 1 \right\rangle +\left| 2 \right\rangle )/\sqrt{2}$.   }
    \label{fig:fig12}
\end{figure}

\textcolor{black}{
 Figure \ref{fig:fig12} presents the dynamics in the excited states at 10 K and 298 K with an initial condition in the upper excited state $\left| 2 \right\rangle$ (panel (a)) or in a superposition of the two excited states $(\left| 1 \right\rangle +\left| 2 \right\rangle )/\sqrt{2}$ (panel (b)). Table \ref{tab:two_bath_modes} gives the number of modes used in HEOM or T-TEDOPA simulations and the corresponding number of complex elements. The HEOM level and the MPS parameters $r$ and $d$ are gathered in the caption. (i) \textit{Continuous case}. The MPS format reveals its efficiency in HEOM since it allows the computation with a maximum tensor-train rank $r_{max} = 60$ while in the standard storage at the level $L=8$, the TM\&FIT and the FP-HEOM applications should need a prohibitive number of complex matrix elements, as shown in Table \ref{tab:two_bath_modes}. T-TEDOPA confirms its performance by requiring fewer resources as expected. It succeeds in reaching room temperature quite easily by increasing the number of modes by 50\%. 
(ii) \textit{LVC case}.  The simulation with all the discrete modes of the LVC model succeeds with the T-TEDOPA method only. We have not been able to carry it out with D-HEOM in this discrete case. Unlike the pure dephasing example, numerical instabilities occur in the D-HEOM propagation in this strongly coupled system, requiring $L = 8$.} The early dynamics with the molecular LVC data merges the predictions of the continuous model for about forty femtoseconds, which corresponds to two oscillation periods of the main high frequency modes. Temperature has very few effects in this simulation. One observes more damping at 298 K but the average populations are quasi similar. The temperature effect is more obvious in the continuous case.

\begin{table*}[ht]
    \centering
    \textcolor{black}{
    \begin{tabular}{lcccccc}
        \toprule
        \toprule
        & \multicolumn{5}{c}{Method} \\
        \cmidrule{3-7}
        T & $J(\omega)$ & ~ & TM\&FIT-HEOM & FP-HEOM  & T-TEDOPA & T-TEDOPA$_{LVC}$ \\
        \midrule
        10 K & $J_d$  & ~ & ~ & \\
         ~  & ~ & $K$ & 6 + 2 & 14 & 60  & 35 \\
         ~ & $J_{od}$  & ~ & ~ & \\
        ~  & ~ & $K$  & 6 + 2 & 14 & 60 & 34 \\
         ~  & ~ & $N^{st}_{HEOM}$ & 2,941,884 & 121,041,360 & ~  & ~ \\
        ~  & ~ & $N^{MPS}_{HEOM}$ & 486,780 & 875,580 & ~  & ~ \\
         ~  & ~ & $N_{TEDOPA}$ & ~ & ~ & 76,800 & 276,000 \\
         \midrule
        298 K & $J_d$  & ~ & ~ & \\ 
        ~  & ~ &$K$ & 6 + 2 & 14 & 90 & 70 \\
        ~ & $J_{od}$  & ~ & ~ & \\
        ~  & ~ & $K$ & 6 + 2 & 14 & 90 & 68 \\
         ~  & ~ & $N^{st}_{HEOM}$ & 2,941,884 & 121,041,360 & ~  & ~ \\
        ~  & ~ & $N^{MPS}_{HEOM}$ & 486,780 & 875,580 & ~  & ~ \\
         ~  & ~ & $N_{TEDOPA}$ & ~ & ~ & 115,200 & 552,000 \\
        \bottomrule
        \bottomrule 
    \end{tabular}}
    \caption{\textcolor{black}{ \label{tab:two_bath_modes} Number of modes for each bath of the two-bath model given by Eqs.(\ref{eq:MTMFIT}) for TM$\&$FIT-HEOM, (\ref{eq:MmodesFP}) for FP-HEOM and (\ref{eq:MmodeTEDOPA}) for T-TEDOPA. The total number $K$ is the sum of the modes of each bath for estimating the number of complex elements given by Eqs.(\ref{eq:NHEOMst}) (without ADOs filtering) and (\ref{eq:NHEOMMPS}) for TM$\&$FIT-HEOM and FP-HEOM, (\ref{eq:NTEDOPA}) for T-TEDOPA. For HEOM, the level is $L = 8$, $r_{max} = 60$ (it should be noted that $r$ may be smaller than $r_{max}$ during the propagation) and the tolerance is of $10^{-10}$. For T-TEDOPA with sampling of the continuous densities, $r=8$ and $d=10$. The cutoff is -0.01 Hartree at 298 K and 0 Hartree at 10 K. For the LVC case, $r=20$ and $d=10$. Only the positive frequencies are used at 10 K since the Bose function is negligible. All the corresponding frequencies on the negative axis are added at 298 K.  } } 
\end{table*}

\begin{figure}
    \centering
    \includegraphics [width =1.\columnwidth]{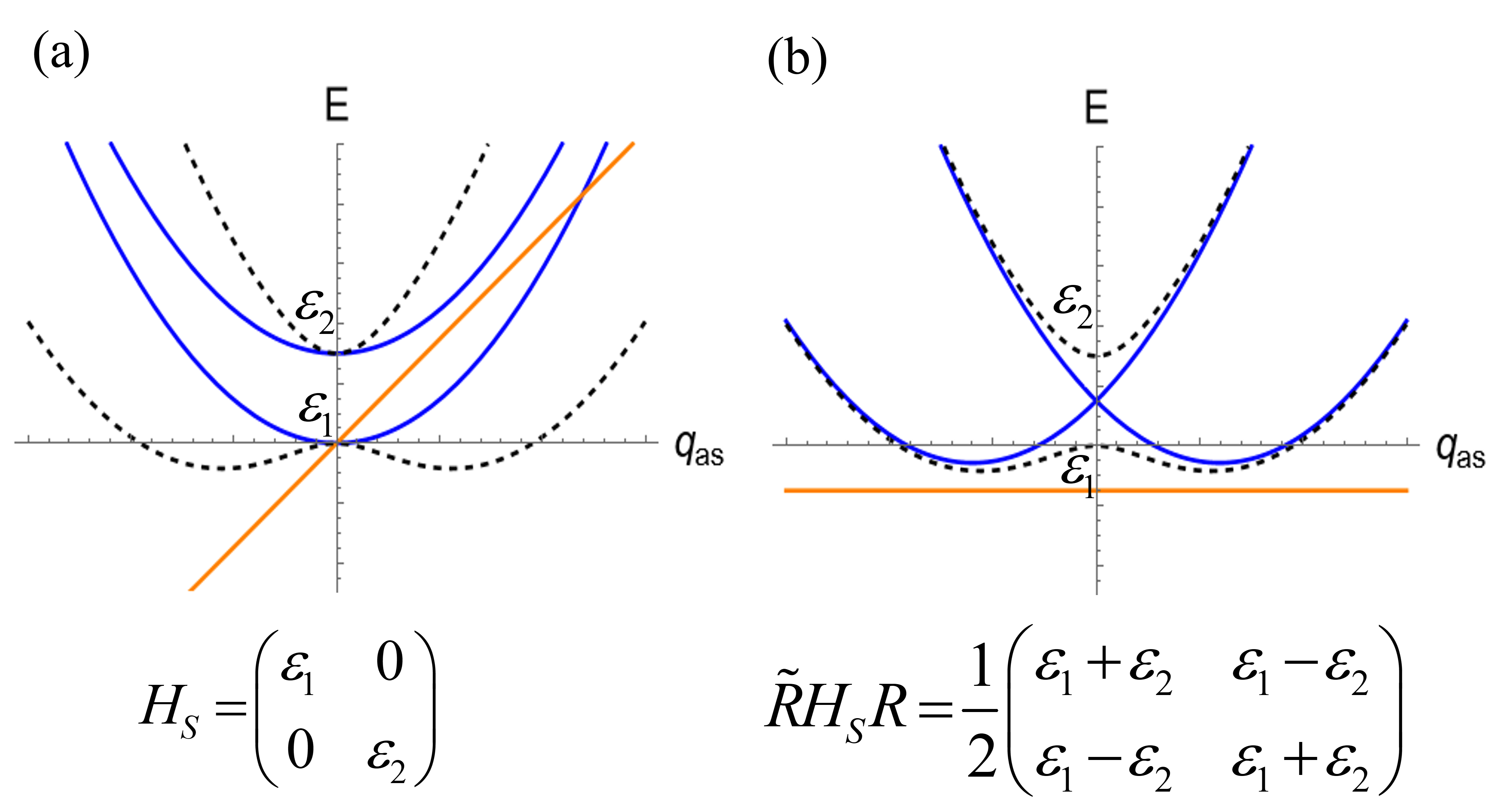}
    \caption{1D cut in diabatic (solid lines) and adiabatic (dashed lines) potential energy surfaces and interstate coupling along an antisymmetric mode $q_{as}$. (a) Diabatic basis in which the interstate coupling is antisymmetric and vanishes in the electronic system Hamiltonian $H_S$ at the reference Franck-Condon position; (b) Diabatic basis after transformation by a rotation matrix of $\pi/4$.   }
    \label{fig:fig13}
\end{figure}

Dynamics shown in Fig.\ref{fig:fig12} deserves some comments. The evolution of the populations seems very similar after the early ultra fast decay of the upper excited state leading to a transitory equal population in both excited states (panel (a)) as in the case of an initial superposition with equal weights (panel (b)). The weak oscillation pattern is the same and the population becomes almost constant after about 150 fs. However, with the initial condition  $\left| 2 \right\rangle$ there is no coherence between the two states during the process while the electronic coherence of the initial superposition persists for some time in the second case.  This leads to a different behavior when the electronic system is analyzed in another diabatic representation. Figure \ref{fig:fig13}(a) schematizes the initial diabatic representation used for the propagation by showing in solid lines a 1D cut in the potential energy surfaces and in the interstate coupling along an antisymmetric mode $q_a$ for an arbitrary value of the symmetrical modes that does not coincide with the conical intersection geometry ($\epsilon_1 \neq \epsilon_2$). The electronic diabatic basis being defined at an arbitrary unitary transformation, the orthogonal transformation by a rotation matrix $R$ of $\pi/4$, one obtains the interesting representation displayed in Fig.\ref{fig:fig13}(b) in terms of two right-left wells \cite{Ho2019,Lasorne2023}. The electronic interstate coupling that is antisymmetric in the first basis becomes symmetrical in the new one. Obviously both representations leads to the same adiabatic basis represented in dashed lines in Fig.\ref{fig:fig13}(a) and (b). Working in this new representation also requires the transformation of the system-bath coupling $H_{SB}$.  

\begin{figure}
    \centering
    \includegraphics [width =0.9\columnwidth]{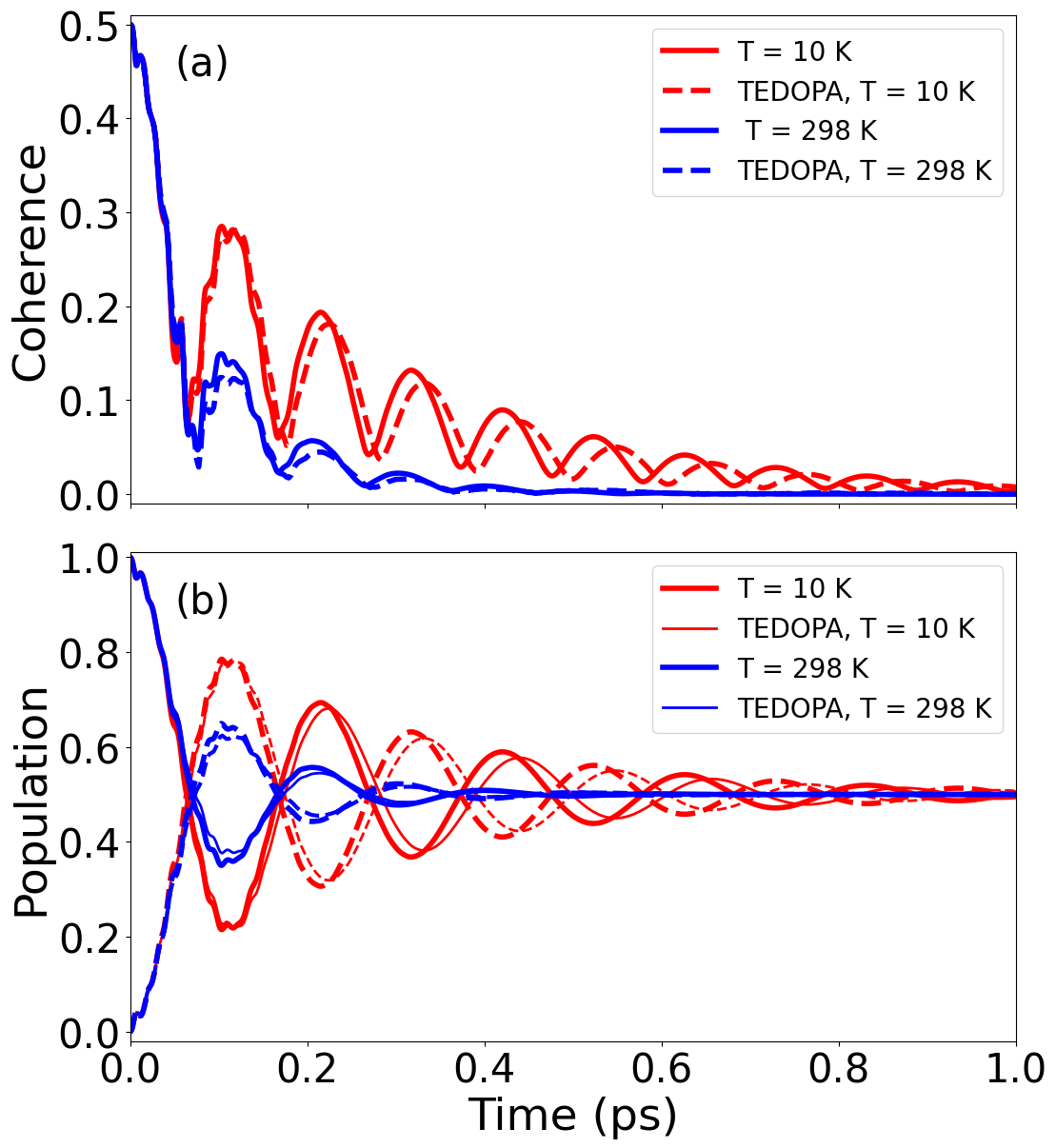}
     \caption{(a) Modulus of the coherence $|\rho_{S_{12}}(t)|$ between the excited states when the system is prepared in a superposed state $(\left| 1 \right\rangle +\left| 2 \right\rangle )/\sqrt{2}$ at 10 K and 298 K computed by HEOM or T-TEDOPA. (b) Population in the $\left| R \right\rangle$ and  $\left| L \right\rangle$ electronic states of the rotated diabatic representation. The HEOM and the T-TEDOPA results are in thick lines and thin lines respectively. In each case, the population in $\left| R \right\rangle$ or  $\left| L \right\rangle$  are in solid or dashed lines.    }
    \label{fig:fig14}
\end{figure}

The initial superposed state $(\left| 1 \right\rangle +\left| 2 \right\rangle )/\sqrt{2}$ corresponds to the preparation in one or the other well $\left| L \right\rangle$ or $\left| R \right\rangle$ according to the sign of the superposition. Figure \ref{fig:fig14}(a) gives the evolution of the coherence $\left| {{\rho }_{{{S}_{12}}}} \right|$ of the initial superposition $(\left| 1 \right\rangle +\left| 2 \right\rangle )/\sqrt{2}$. The electronic coherence persists during 1 ps at 10 K and 350 fs at 298 K. Figure \ref{fig:fig14}(b) presents the population in the two $\left| R \right\rangle$ or $\left| L \right\rangle$ states of the rotated basis. The damped Rabi oscillation between these two states persists as long as the coherence survives. The slight dephasing between the results in HEOM and T-TEDOPA may be due to a lack of convergence in HEOM at long time due to a too small level (level $L=8$ in HEOM versus a Fock space dimension $d=10$ in T-TEDOPA).

\section{Summary and concluding remarks}
\label{sec:conclu}
HEOM and T-TEDOPA are advanced simulation methods for treating non-Markovian open quantum systems with structured environments. This challenging regime is often encountered in simulations of models calibrated from molecular dynamics \cite{Mangaud2015,Geva2020,DunnettMB2021} or LVC $\textit{ab}$ $\textit{initio}$ models in molecular systems \cite{Breuil2021,Schröder2019,LiuShi2024}, and in many domains of quantum technology. In this work, our goal is to bring together different algorithms to tackle the temperature problem in open quantum systems in strong interaction with structured environments. The central point that we want to illustrate is the effort in the opposite direction to decrease or increase the temperature depending on whether the dynamics is treated by HEOM or by the multi-dimensional wave packet approach T-TEDOPA. \textcolor{black}{ We illustrate different ways of choosing the artificial modes that are the essential tool of the expansion of the two-time bath correlation function $C(t)$ in HEOM. Directly fitting the complex correlation function in terms of exponential functions is possible by different algorithms \cite{Wu2015,Nakamura2018,Lambert2019,Kleinekathofer2019,Ikeda2020,Plenio2020,Yan2022,Lambert2023,ThossBorrelli2024}. However, common methods start from the spectral density $J(\omega)$ or the Fourier transform of $C(t)$ that is the temperature dependent spectral density $J_{\beta}(\omega)$, in particular in FP-HEOM or in T-TEDOPA. The two illustrating examples differ by the shape of the spectral density and by the strength of the system-environment coupling for a given timescale. This brings qualitative insight on the resources required in each method. However, this cannot be generalized easily. Further investigation is necessary to analyze the role of the simulation duration and of the spectral density complexity.      }

\textcolor{black}{In the HEOM method, the three investigated approaches have strong and weak points depending on the shape of the spectral density and on the coupling strength. The parametrization of $J(\omega)$ by Lorentzian functions is interesting only when there are few emerging peaks linked to some underdamped strongly coupled vibrational modes. This is frequently the case in organic molecules. For simulations at low temperature, the drawback posed by the poles of the Bose function may be effectively addressed by fitting their contribution to the bath correlation function. The effective number of modes and then the resources involved in the TM$\&$FIT strategy can remain competitive.  The very interesting advance based on the fit of $J_{\beta}(\omega)$ by the barycentric method (FP-HEOM) bypasses the problem of Matsubara terms and is  undoubtedly a promising method by involving a small number of artificial modes. In the examples covered here, TM$\&$FIT demands a little less resources than the FP-HEOM strategy. It is likely that FP-HEOM may become more efficient for more structured densities or in super-Ohmic cases as in solid phase. Then, in the extension of the TM parametrization with super-Ohmic Lorentzian functions, each function involves four-poles in the upper complex plane Lorentzian instead of two, which doubles the number of modes \cite{MangaudMeier2017}. The D-HEOM strategy based on the sampling of $J(\omega)$ with undamped modes has proven to be ineffective in our strong coupling example implying high hierarchy levels.}

\textcolor{black}{In both illustrations presented here, the T-TEDOPA method is the most efficient with respect to the storage resources and the computational time even when the number of modes is larger than in some HEOM strategies. Furthermore, the method may bring a more detailed information on the bath dynamics by transforming back from the chain model to the star model. HEOM gives more global statistical information about the bath distribution \cite{Shi2012,Shi2014,ChinChevet2019,MDL_19,Jaouadi2022}. The T-TEDOPA algorithm may benefit from an asymmetrical frequency range for the sampling. The truncation of the correlation function after a typical timescale of the dynamics before the Fourier transforming coupled with the chain mapping seems to be an outperforming strategy of T-TEDOPA \cite{Plenio2024,deVega-Wolf2015}. Furthermore, the shorter chain length for a higher temperature in the Ohmic-like spectral density pure dephasing example brings interest and we hope to investigate it more deeply in future work.}   
 
\section*{Supplemental Material}
The Supplementary Material gives the detailed HEOM equations with the different artificial modes (TM, FP-HEOM and D-HEOM). It gathers the parameters of the TM functions fitting the different spectral densities and all the parameters of the exponential functions fitting the contribution of the Matsubara terms to the bath correlation functions at low temperature. We also illustrate the convergence of the number of undamped discrete modes for the D-HEOM and T-TEDOPA simulations in the pure dephasing qubit. 

\section{Acknowledgements}
We warmly acknowledge Dr B. Lasorne and J. Galiana (PhD) for the $\textit{ab}$ $\textit{initio}$ data used to calibrate the two-bath model. B.LD acknowledges the support of iSiM Sorbonne. EM acknowledges support from ANR project NQESim (Grant No. ANR-23-CE29-0024-01). AWC wishes to acknowledge support from ANR Project ACCEPT (Grant No. ANR-19-CE24-0028).

\section*{Data availability}
The data that support this study are available on request from the authors. All the T-TEDOPA simulations were done with the MPSDynamics Julia package \cite{MPSDynamics} that can be found along with the documentation at \url{https://github.com/shareloqs/MPSDynamics}.

\bibliographystyle{unsrt}
\bibliography{bib_temp.bib} 
\end{document}



\title{Supplemental Material: Managing Temperature in Strongly Interacting Quantum Systems Coupled with Structured Environments}



\author{Brieuc Le D\'e}
 \affiliation{Institut des Nanosciences de Paris, Sorbonne Université, CNRS, F-75005 Paris, France} 

 \author{Amine Jaouadi}
 \affiliation{LyRIDS, ECE Paris, Graduate School of Engineering, Paris, F-75015, France}

\author{Etienne Mangaud}
 \affiliation{MSME, Universit\'e Gustave Eiffel, UPEC, CNRS, F-77454 Marne-La-Vall\'ee, France}

\author{Alex W. Chin}
\affiliation{Institut des Nanosciences de Paris, Sorbonne Université, CNRS, F-75005 Paris, France}

\author{Mich\`ele Desouter-Lecomte}
\email{michele.desouter-lecomte@universite-paris-saclay.fr}
\affiliation{Institut de Chimie Physique, Universit\'e Paris-Saclay-CNRS, UMR8000, F-91400 Orsay, France} 


\date{\today}

\begin{abstract}
This Supplemental Material gives the detailed HEOM equations with the different artificial modes (TM-HEOM, FP-HEOM and D-HEOM). It gathers the parameters of the two-pole Lorentzian Tannor-Meier functions fitting the spectral densities used in the simulations. Parameters of the exponential functions fitting the  contribution of the Matsubara terms to the bath correlation functions $C(t)$ are also reported. Finally, we illustrate the convergence of the number of discrete modes used in D-HEOM and T-TEDOPA in the pure dephasing qubit.

\end{abstract}

\maketitle

\section{HEOM with the different artificial modes}
\label{sec:APP_HEOM}
We first summarize the working HEOM coupled equations for the Tannor-Meier expansion of the spectral density (TM-HEOM), the FP-HEOM method and the discrete modes D-HEOM.  We distinguish the contribution of the modes with positive or negative frequencies. The HEOM coupled equations are given for a single bath with a coupling operator $S$. Generalization for more baths involves a supplementary summation over the baths and the use of different coupling operators $S_{\alpha}$ for each bath.

\subsection{Modes from the poles of $J(\omega)$}
When the spectral density is fitted by a sum of Tannor-Meier Lozentzian functions \cite{Meier1999,Pomyalov2010}
\begin{equation}
 J^{TM}\left( \omega  \right)  \approx\sum\limits_{l=1}^{{{N}_{l}}}{\frac{{{p}_{l}}\omega }{\left[ {{\left( \omega +{{\Omega }_{l}} \right)}^{2}}+\Gamma _{l}^{2} \right]\left[ {{\left( \omega -{{\Omega }_{l}} \right)}^{2}}+\Gamma _{l}^{2} \right]}}
\label{eq:JTM}
\end{equation}
the relevant poles in the upper half-plane associated with each Lorentzian function are
\begin{align}  \label{eq:gammaTM}
 {{\gamma_{l1}=\Omega }_{l}} +i{{\Gamma }_{l}} \nonumber \\ 
 {{\gamma_{l2}=-\Omega }_{l}}+i{{\Gamma }_{l}}.
 \end{align} 
We distinguish the contributions of the modes with rates $\gamma_{l1}$, $\gamma_{l2}$ [Eq.(\ref{eq:gammaTM})] coming from the $N_l$ Lorentzian functions and the $M_a$ Matsubara modes that are given by the poles on the imaginary axis ($\omega=0$). The Matsubara rates are 
\begin{equation}
\gamma_{Mk}=i2\pi k/\beta.    
\end{equation}
The $\alpha_k$ complex coefficients associated to the poles of the Lorentzian functions are
\begin{align}  \label{eq:alphaTM}
  & {{\alpha }_{l1}}=\frac{{{p}_{l}}}{8{{\Omega }_{l}}{{\Gamma }_{l}}}\left[ \coth \left( \frac{\beta }{2}({{\Omega }_{l}}+i{{\Gamma }_{l}}) \right)-1 \right] \nonumber \\  
 & {{\alpha }_{l2}}=\frac{{{p}_{l}}}{8{{\Omega }_{l}}{{\Gamma }_{l}}}\left[ \coth \left( \frac{\beta }{2}({{\Omega }_{l}}-i{{\Gamma }_{l}}) \right)+1 \right]. 
\end{align} 
 The parameters for the Matsubara terms are 
 \begin{equation}
 \alpha_{Mk}=2iJ(\gamma_{Mk}/\beta.    
 \end{equation}
 In principle $M_{\alpha}$ is infinite. The matrices are denoted with a triple multi-index ${{\rho }_{\mathbf{m,n},\mathbf{M}}}$ giving the occupation number in the modes with $\gamma_{l1}$, $\gamma_{l2}$ and the Matsubara modes with $\gamma_{Mk}$ respectively. The HEOM are: 
\begin{align} \label{eq :HEOMTM}
  & {{{\dot{\rho }}}_{\mathbf{m,n},\mathbf{M}}}=-i\left[ {{H}_{S}},{{\rho }_{\mathbf{m,n},\mathbf{M}}} \right] \nonumber \\ 
 & +i\left\{ \sum\limits_{l=1}^{{{N}_{l}}}{\left( {{m}_{l}}{{\gamma }_{l1}}+{{n}_{l}}{{\gamma }_{l2}} \right)+\sum\limits_{k}^{M}{{{M}_{k}}{{\gamma }_{Mk}}}} \right\}{{\rho }_{\mathbf{m,n},\mathbf{M}}} \nonumber \\ 
 & -i\sum\limits_{l=1}^{{{N}_{l}}}{\left[ S,\left( {{\rho }_{\mathbf{m}_{l}^{+}\mathbf{,n},\mathbf{M}}}+{{\rho }_{\mathbf{m,n}_{l}^{+},\mathbf{M}}} \right) \right]} \nonumber  \\ 
 & -i\sum\limits_{k=1}^{M_a}{\left[ S,{{\rho }_{\mathbf{m,n},\mathbf{M}_{k}^{+}}} \right]} \nonumber \\ 
 & -i\sum\limits_{l=1}^{{{N}_{l}}}{{{m}_{l}}\left( {{\alpha }_{l2}}S{{\rho }_{\mathbf{m}_{l}^{-}\mathbf{,n},\mathbf{M}}}-{{{\bar{\alpha }}}_{l1}}{{\rho }_{\mathbf{m}_{l}^{-}\mathbf{,n},\mathbf{M}}}S \right)}  \nonumber \\ 
 & -i\sum\limits_{l=1}^{{{N}_{l}}}{{{n}_{l}}\left( {{\alpha }_{l1}}S{{\rho }_{\mathbf{m,n}_{l}^{-},\mathbf{M}}}-{{{\bar{\alpha }}}_{l2}}{{\rho }_{\mathbf{m,n}_{l}^{-},\mathbf{M}}}S \right)} \nonumber \\ 
 & -i\sum\limits_{k=1}^{M_a}{{{M}_{k}}{{\alpha }_{Mk}}\left[ S,{{\rho }_{\mathbf{m,n},\mathbf{M}_{k}^{-}}} \right]} 
\end{align}
where the upper bar means the complex conjugate. 

\subsection{Modes from the poles of $J_{\beta}(\omega)$}
Each auxiliary matrix is labelled by the collective index $\mathbf{m,n}$ where $\mathbf{m}$ is the index giving the occupation number in the $N_{P}$ selected poles with parameters $\gamma_{Pk1} = \Omega_k + i \Gamma_k$ and $\mathbf{n}$ refers to the same number of modes with $\gamma_{Pk2} = -\Omega_k + i \Gamma_k$. The coefficients $\alpha_{Pk1}$ and and $\alpha_{Pk2}$ are given by  \begin{align}
  & {{\alpha }_{Pk1}} = 2i Res[P_k]  \nonumber \\
  & {\alpha }_{Pk2} = {\bar{\alpha }}_{Pk1}
  \label{eq:alphaFP}
 \end{align}
 where $Res[P_k]$ is the residue at each selected pole. The equations are derived in the supplemental material of Ref.\cite{Xu2022} using scaled density matrices ${{\tilde{\rho }}_{\mathbf{m,n}}}=\prod\limits_{k}{\sqrt{{{m}_{k}}!\alpha _{Pk1}^{{{m}_{k}}}}}\sqrt{{{n}_{k}}!\alpha _{Pk2}^{{{n}_{k}}}}{{\rho }_{\mathbf{m,n}}}$. 

\begin{align}  \label{eq :HEOMFP}
  & {{{\dot{\tilde{\rho }}}}_{\mathbf{m,n}}}=-i\left[ {{H}_{S}},{{{\tilde{\rho }}}_{\mathbf{m,n}}} \right] \nonumber  \\ 
 & +i\sum\limits_{k=1}^{{{N}_{FP}}}{\left( {{m}_{k}}{{\gamma }_{Pk1}}+{{n}_{k}}{{\gamma }_{Pk2}} \right)}{{{\tilde{\rho }}}_{\mathbf{m,n}}} \nonumber   \\ 
 & -i\sum\limits_{k=1}^{{{N}_{P}}}{\sqrt{\left( {{m}_{k}}+1 \right){{\alpha }_{Pk1}}}\left[ S,{{{\tilde{\rho }}}_{\mathbf{m}_{k}^{+}\mathbf{,n}}} \right]} \nonumber   \\ 
 & -i\sum\limits_{k=1}^{{{N}_{P}}}{\sqrt{\left( {{n}_{k}}+1 \right){{\alpha }_{Pk2}}}\left[ S,{{{\tilde{\rho }}}_{\mathbf{m,n}_{k}^{+}}} \right]}  \nonumber  \\ 
 & -i\sum\limits_{k=1}^{{{N}_{P}}}{\sqrt{{{m}_{k}}{{\alpha }_{Pk1}}}S{{{\tilde{\rho }}}_{\mathbf{m}_{k}^{-}\mathbf{,n}}}} \nonumber   \\ 
 & -i\sum\limits_{k=1}^{{{N}_{P}}}{\sqrt{{{n}_{k}}{{\alpha }_{Pk2}}}{{{\tilde{\rho }}}_{\mathbf{m,n}_{k}^{-}}}}S  
\end{align}

\subsection{Discrete modes by sampling $J(\omega)$}
Each auxiliary matrix is labelled by the collective index $\mathbf{m,n}$ where $\mathbf{m} = (m_1,.,m_k,..,m_M)$  and $\mathbf{n} = (n_1,.,n_k,..,n_M)$ give the occupation numbers in the $M$ modes with factors ($e^{i\omega_k t}$) and ($e^{-i\omega_k t}$) respectively. The HEOM without scaling the auxiliary matrices are written: 
\begin{align}  \label{eq:HEOMDM}
  & {{{\dot{\rho }}}_{\mathbf{m,n}}}=-i\left[ {{H}_{S}},{{\rho }_{\mathbf{m,n}}} \right]+i\sum\limits_{k=1}^{M}{\left( {{m}_{k}\gamma_{k1}}+{{n}_{k}\gamma_{k2}} \right)}{{\rho }_{\mathbf{m,n}}}  \nonumber \\
 & -i\sum\limits_{k=1}^{M}{\left[ S,\left( {{\rho }_{{{\mathbf{m}}_k^{\mathbf{+}}}\mathbf{,n}}}+{{\rho }_{\mathbf{m,}{{\mathbf{n}}_k^{\mathbf{+}}}}} \right) \right]}  \nonumber \\
 & -i \sum\limits_{k=1}^{M}{m_k\left( {{\alpha }_{k1}}S{{\rho }_{{{\mathbf{m}}_k^{-}}\mathbf{,n}}}-{{\alpha }_{k2}}{{\rho }_{{{\mathbf{m}}_k^{-}}\mathbf{,n}}}S \right)}   \nonumber  \\
 & -i \sum\limits_{k=1}^{M}{n_k\left( {{\alpha }_{k2}}S{{\rho }_{\mathbf{m,}{{\mathbf{n}}_k^{-}}}}-{{\alpha }_{k1}}{{\rho }_{\mathbf{m,}{{\mathbf{n}}_k^{-}}}}S \right)}] 
\end{align}
where $\mathbf{m}_{k}^{\pm }=\{{{m}_{1,}}...,{{m}_{k}}\pm 1,..{{m}_{M}}\}$ with the same notation for $\mathbf{n}_{k}^{\pm }$. The $\gamma_{k}$ are given \begin{align}  \label{eq:gammaDM}
 {{\gamma }_{m1}}={{\omega }_{m}}  \nonumber  \\
 {{\gamma }_{m2}}=-{{\omega }_{m}}.
\end{align}
The corresponding coefficients are:
\begin{align}
\label{eq:alphaMD}
\alpha _{m1}^{{}}=\frac{c_{m}^{2}}{2{{\omega }_{m}}}\left( \frac{1}{{{e}^{\beta {{\omega }_{m}}}}-1} \right)  \nonumber  \\
{\alpha_{m2}}=\frac{c_{m}^{2}}{2{{\omega }_{m}}} \left( \frac{{{e}^{\beta {{\omega }_{m}}}}}{{{e}^{\beta {{\omega }_{m}}}}-1} \right).
\end{align}

\section{SPECTRAL DENSITY PARAMETERS}
The parameters $p_{l}$, $\Omega_{l}$ and $\Gamma_{l}$ of the Tannor-Meier Lorentzian functions [Eq.(\ref{eq:JTM})] used for the spectral densities $J_{1}(\omega)$ and $J_{2}(\omega)$ of the pure dephasing spin-Boson model are given in Table \ref{tab:table_paramters_J}. 
\begin{table}[ht]
    \begin{tabular}{cccc}
    \\
    \hline
    \hline
$J_{1}$ or $J_{2}$ & $p_{l}$ (a.u) & $\Omega_{l}$ (a.u) & $\Gamma_{l}$ (a.u) \\ 
 \hline
$J_{1}$ & $2 \times 10^{-9}$ & $6 \times 10^{-3}$ & $5 \times 10^{-3}$ \\
$J_{2}$ & $2.2 \times 10^{-10}$ & $6 \times 10^{-3}$ & $1 \times 10^{-3}$ \\
\hline
\hline
    \end{tabular}
\caption{\label{tab:table_paramters_J} Parameters of the Tannor-Meier Lorentzian functions used for the spectral densities of the pure dephasing Spin-Boson model.}
\end{table}

The parameters used in the two-bath model for fitting the spectral density of the tuning bath $J_{d}(\omega)$ and the coupling bath $J_{od}(\omega)$ are gathered in Table \ref{tab:table_twobaths}.

\begin{table}[ht]
    \begin{tabular}{cccc}
    \hline
$J_{d}$ or $J_{od}$ & $p_{l}$ (a.u) & $\Omega_{l}$ (a.u) & $\Gamma_{l}$ (a.u) \\ 
 \hline
 \hline
$J_{d}$ & $5.30 \times 10^{-10}$ & $1.078687 \times 10^{-2}$ & $3.40 \times 10^{-4}$ \\
~      & $1.06 \times 10^{-10}$ & $7.6634\times 10^{-3}$ & $4.0 \times 10^{-4}$ \\
 ~      & $5.30\times 10^{-11}$ & $5.4517 \times 10^{-3}$ & $3.40 \times 10^{-4}$ \\
$J_{od}$ & $3.82 \times 10^{-10}$ & $1.078275 \times 10^{-2}$ & $3.40 \times 10^{-4}$ \\
 ~       & $1.45 \times 10^{-10}$ & $7.6634 \times 10^{-3}$ & $4.0 \times 10^{-4}$  \\
 ~       & $4.58 \times 10^{-11}$ & $5.286 \times 10^{-3}$ & $4.50 \times 10^{-4}$ \\
\hline
\hline
    \end{tabular}
\caption{\label{tab:table_twobaths} Parameters of the Tannor-Meier Lorentzian functions fitting the spectral densities of the two-bath model. $J_{d}(\omega)$ corresponds to the tuning bath making fluctuate the energies and $J_{od}(\omega)$ refers to the coupling bath inducing variation of the interstate coupling. }
\end{table}

\section{Parameters of the exponential functions fitting  of the Matsubara contribution to $C(t)$}
The contribution of the Matsubara terms $C_{M}(t)$ to the bath correlation function $C(t)$ is fitted by a sum of exponential functions as the following: 

\begin{equation}
    f(t)=\sum\limits_{k=1}^{N}a_{k}\times e^{(-b_{k}*t)}.
\end{equation}

\begin{figure}[!ht]
    \centering
    \includegraphics [width =1.\columnwidth, height = 5 cm]{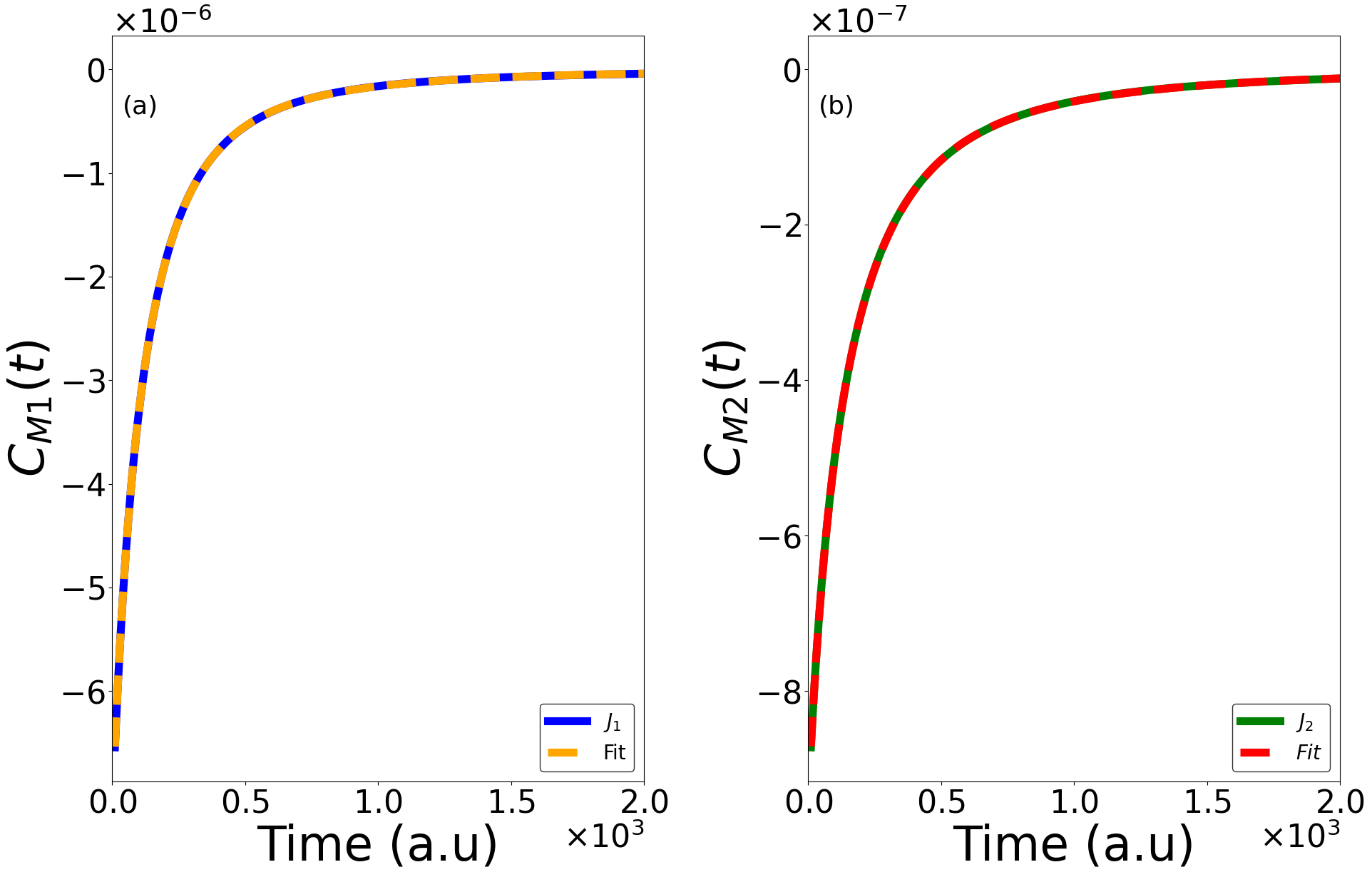}
    \caption{Fit of the Matsubara contributions $C_{M1}(t)$ and $C_{M2}(t)$ related to $J_{1}$ and $J_{2}$ by 4 exponential functions. (a) $C_{M1}(t)$ and (b) $C_{M2}(t)$. }
    \label{fig:FigureSM1}
\end{figure}

In the simulation of the pure dephasing spin-Boson case, we use four exponential functions to fit the Matsubara contributions $C_{M1}(t)$ and $C_{M2}(t)$ to $C_{1}(t)$ and $C_{2}(t)$ respectively as shown is Fig.\ref{fig:FigureSM1}. The parameters $a_{k}$ and $b_{k}$ are given in Table \ref{tab:table_fitting_J1J2}.

\begin{table*}[!ht]
   \centering
    \begin{tabular}{lccccc}
    \hline
    \hline
        ~ & ~ & function 1  & function 2  & function 3  & function 4 
        \\ \hline
        $C_{M1}$  & a & 2.18171e-06 & 3.88737e-06 & 1.00599e-06 & 7.4179e-08 \\ 
        ~ & b & 0.017384 & 0.00643924 & 0.00219971 & 0.000482092  \\
        \\ 
        $C_{M2}$ & a & 4.86192e-07 & 2.73089e-07 & 1.6411e-07 & 1.46568e-08\\ 
        ~ & b & 0.00522371 & 0.0155586 & 0.00174184 & 0.000401763       
        \\ \hline 
        $C_{M1}$ RMS &  6.4082121066e-10 & ~ \\ 
        $C_{M2}$ RMS & 1.18772056479e-10 &  ~ \\ 
        \hline
        \hline
    \end{tabular}
    \caption{\label{tab:table_fitting_J1J2} Fitting parameters of the Matsubara contribution $C_{M1}(t)$ and $C_{M2}(t)$ related to the spectral densities $J_1$ and $J_2$ of the pure dephasing example by 4 exponential functions.}
\end{table*}

\begin{figure}[!ht]
    \centering
    \includegraphics [width =1.\columnwidth, height = 5 cm]{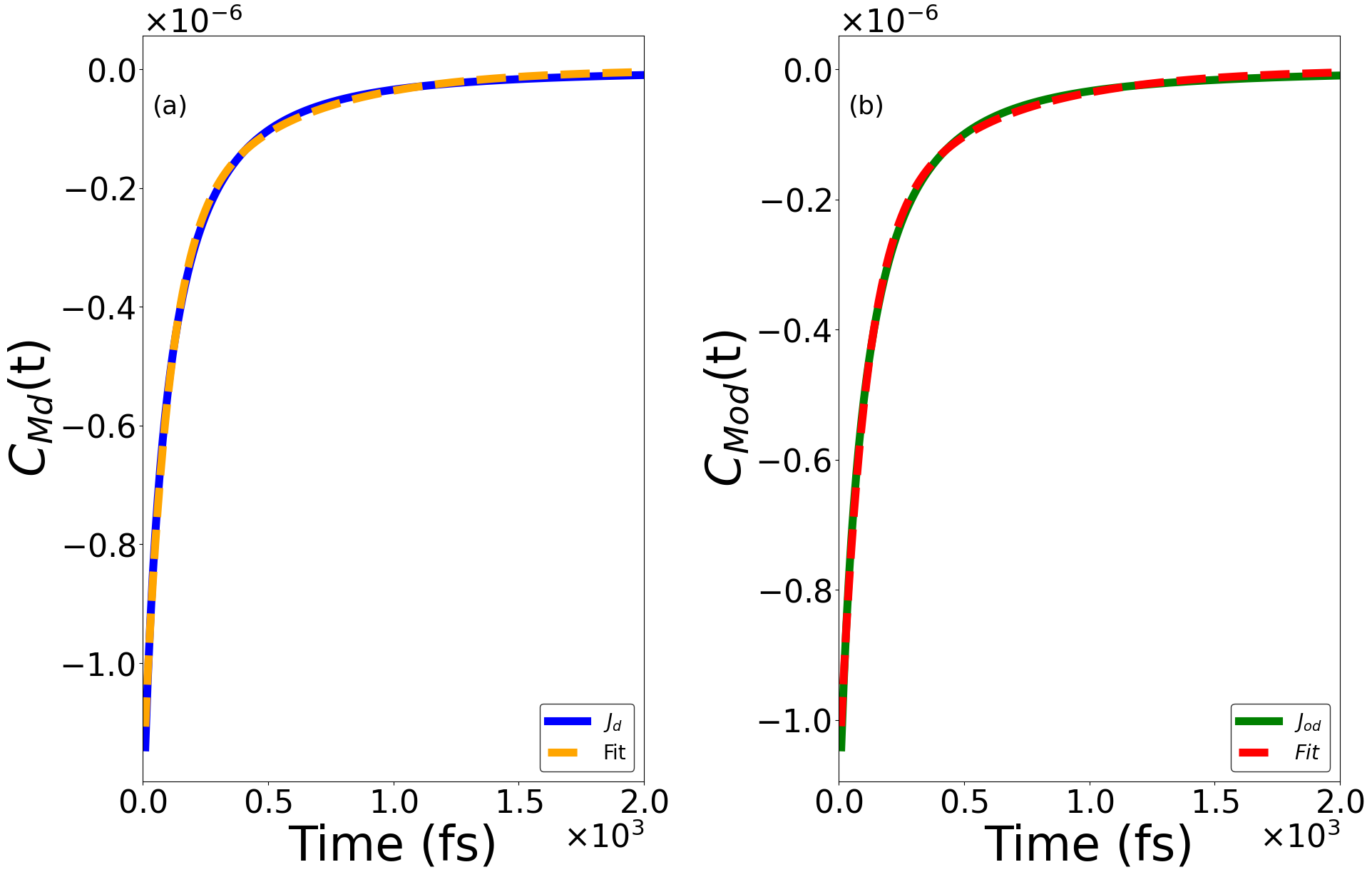}
    \caption{Fit of the Matsubara contributions to the correlation functions of the two-bath model by two exponential functions. (a) $C_{Md}$ (tuning bath) and (b) $C_{Mod}$ (coupling bath). }
    \label{fig:FigureSM2}
\end{figure}

In the two-bath model, the Matsubura contributions to the bath correlation functions related to $J_{d}(\omega)$ and $J_{od}(\omega)$ denoted $C_{Md}(t)$ and $C_{Mod}(t)$ are fitted by two exponential functions. The fitting parameters  $a_{k}$ and $b_{k}$ are given in Table \ref{tab:table_fitting_matsu}. $C_{Md}$ and $C_{Mod}$ are shown in Fig. \ref{fig:FigureSM2} in panels (a) and (b) respectively. 

\begin{table}[!ht]
    \centering
    \begin{tabular}{lccc}
    \hline
    \hline
        ~ & ~ & function 1  & function 2  
        \\ \hline
        $C_{Md}$  & a & -8.25349e-07 & -2.69966e-07 \\ 
        ~ & b & 0.0102391 & 0.00202244 
        \\ 
        $C_{Mod}$ & a & -9.1026e-07 & -2.95188e-07 \\ 
        ~ & b & 0.0106902 & 0.00210567       
        \\ \hline 
        $C_{Md}$ RMS &  2.2977856e-09 & ~ \\ 
        $C_{Mod}$ RMS & 2.1371695e-09 &  ~ \\ 
        \hline
        \hline
    \end{tabular}
    \caption{\label{tab:table_fitting_matsu} Fitting parameters of the Matsubara contributions $C_{Md}$ and $C_{Mod}$ related to the spectral densities $J_d$ and $J_{od}$ of the two-bath model by 2 exponential functions.}
\end{table}

\section{Minimum number of discrete Modes for HEOM and T-TEDOPA in the pure dephasing qubit}

\begin{figure}[!ht]
    \centering
    \includegraphics [width =1.\columnwidth, height = 5 cm]{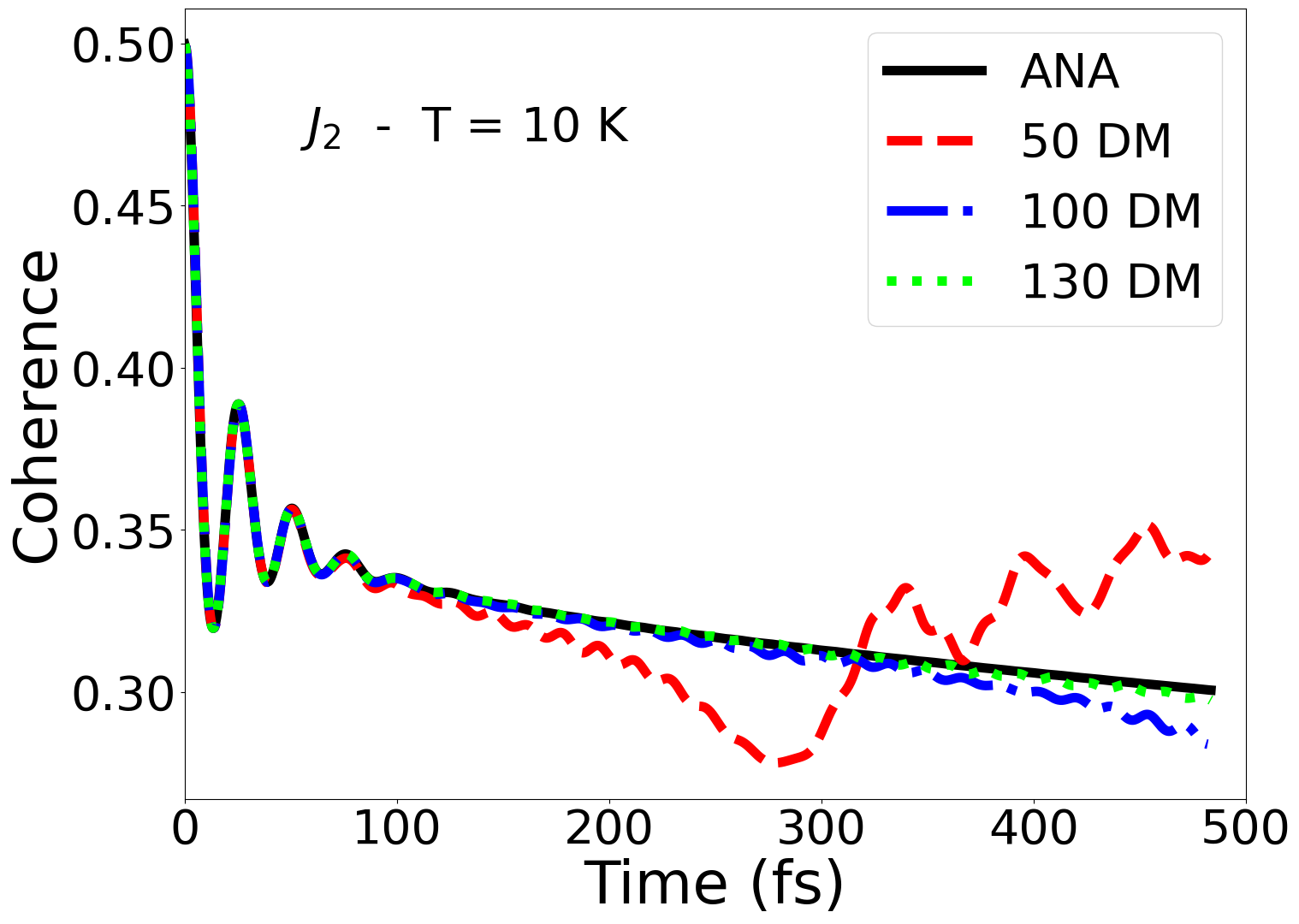}
    \caption{Convergence of the coherence as function of the number $M$ of selected discrete modes with positive frequencies in the case of $J_2(\omega)$ at T = 10 K, used in the simulation of the pure dephasing qubit (the computation used $2M$ modes by adding the symmetrical negative frequencies).}
    \label{fig:FigureSM3}
\end{figure}

In HEOM simulation using undamped discrete modes, we ascertain the minimum number $M$ of selected modes with positive frequencies required to achieve convergence. In practice, the computation involves modes with positive and negative frequencies, i.e. $2M$ modes. Notably, in the HEOM case, the number of modes is higher for the more non-Markovian example involving $J_2(\omega)$ used in the simulation of the pure dephasing qubit, and it increases as the temperature decreases. We present in Fig. \ref{fig:FigureSM3} the convergence of the coherence in the case of $J_2(\omega)$ at T = 10 K. One could notice that the convergence is reached with at least 130 discrete modes presented in green dotted line. 

\begin{figure}[!ht]
     \centering
         \centering
         \includegraphics[width=\columnwidth]{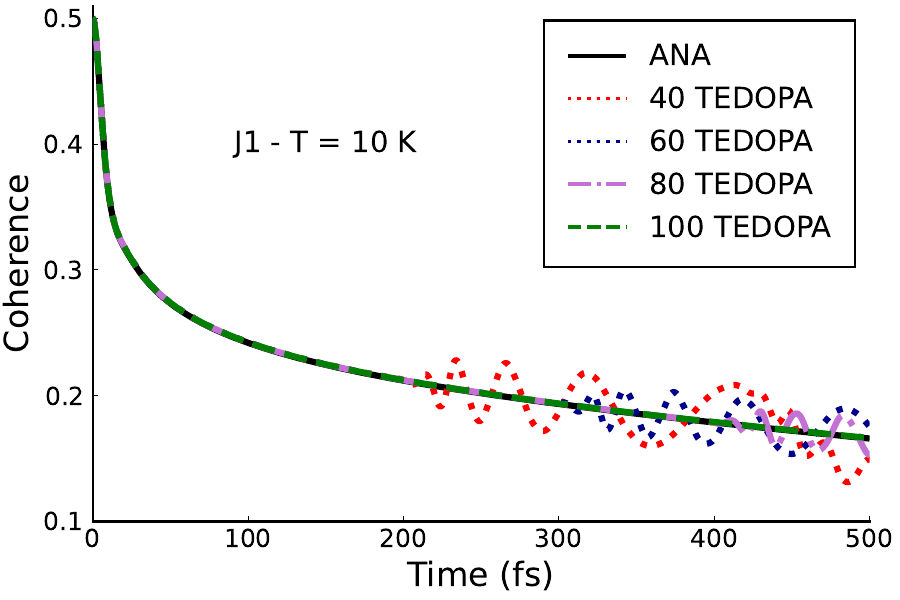}
     \hfill
         \centering      \includegraphics[width=\columnwidth]{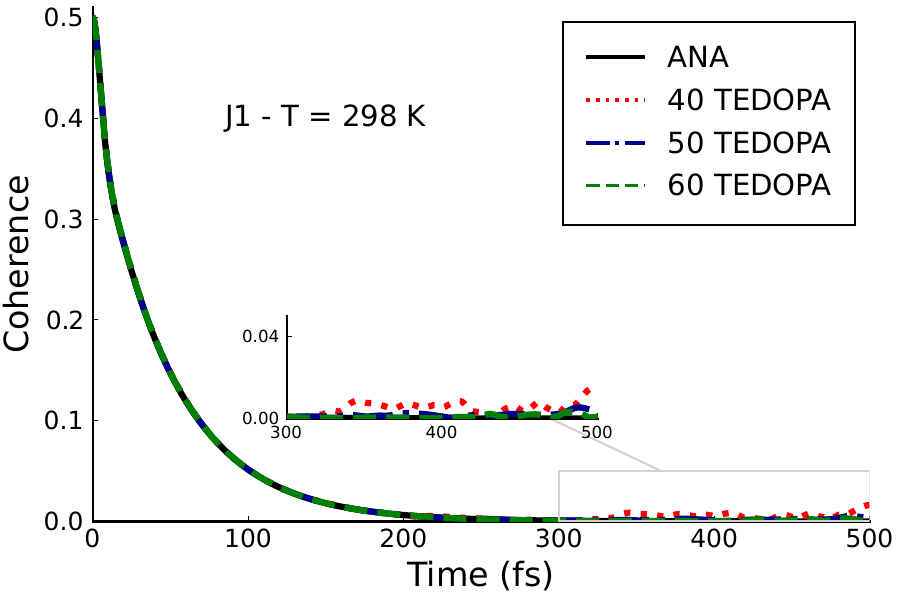}
        \caption{Convergence of the coherence for T-TEDOPA as function of the number $N_{ch}$ of selected discrete modes in the case of $J_1(\omega)$ at T = 10 K (top) and T = 298 K (bottom), used in the simulation of the pure dephasing qubit.}
        \label{fig:FigureSM4}
\end{figure}

Concerning T-TEDOPA simulations, we show the convergence of the number of chain modes $N_{ch}$ for the interesting observation of the spectral density $J_1$ at T = 10 K and T = 298 K (Fig. \ref{fig:FigureSM4}). Convergence is reached at $N_{ch}=100$ at T = 10 K whereas that at T = 298 K, convergence is reached at $N_{ch}=60$. Unlike the HEOM method, $N_{ch}$ is the total number of modes within the chain. 

\newpage

%
%

%


\bibliographystyle{unsrt}
\bibliography{bib_temp.bib}